\documentclass[letterpaper]{iopart}
\usepackage{cite}
\usepackage{amsmath, amssymb}
\usepackage{amsbsy, latexsym, amsfonts, amsthm}
\usepackage{mathtools}
\usepackage{graphicx}
\usepackage{xcolor}
\usepackage{hyperref}
\usepackage[fonts]{faust}

\theoremstyle{plain}
\newtheorem{theorem}{Theorem}
\newtheorem{lemma}{Lemma}
\newcommand{\startproof}{\setlength{\parindent}{0in}\textbf{Proof.} }
\newcommand{\finishproof}{\hfill $\blacksquare$ \\}

\makeatletter

\DeclareFontFamily{U}{bbm}{}											% Load Blackboard Bold
\DeclareFontShape{U}{bbm}{m}{n}{<5><6><7><8><9><10> gen * bbm 
	<10.95> bbm10 <12><14.4> bbm12 <17.28><20.74><24.88> bbm17}{}
\DeclareMathAlphabet{\mathbb}{U}{bbm}{m}{n}

\DeclareFontFamily{U}{rsfs}{}											% Load Formal Script
\DeclareFontShape{U}{rsfs}{m}{n}{<5> rsfs5 <6><7> rsfs7 
	<8><9><10><10.95><12><14.4><17.28><20.74><24.88> rsfs10}{}
\DeclareMathAlphabet{\mathfs}{U}{rsfs}{m}{n}

\newcommand{\dummy}{\rule{0pt}{0pt}}

\newbox\faust@box
\newdimen\faust@dimen

\newtoks\util@toks
\newtoks\defaults@toks

\def\defaults@declare#1#2#3{%
	\util@toks={#1#2{#3}}%
	\edef\@act{\long\def\the\util@toks}\@act%
}

\def\defaults@emptytest#1#2\st@p#3#4{%
	\ifx#1\l@st  \toks@={#3}\relax  \else  \toks@={#4}\relax  \fi%
	\edef\@act{\the\toks@}\@act%
}

\def\defaults@process#1#2{%
	\defaults@checknext{#1}{}{#2}%
}

\def\defaults@checknext#1#2#3{%
	\@ifnextchar[{\defaults@replacenext{#1}{#2}#3\st@p}{#1#2#3}%
}

\def\defaults@replacenext#1#2[#3]#4\st@p[#5]{%
	\IfEmptyArg{#5}{\defaults@checknext{#1}{#2[#3]}{#4}}{\defaults@checknext{#1}{#2[#5]}{#4}}%
}

\newbox\dirac@box
\newtoks\dirac@toks

\makeatother

\DeclareMathOperator\vol{vol}

\newcommand{\Cyl}{\mathrm{Cyl}}

\newcommand{\Hil}{\mfs{H}}

\newcommand{\dilparam}{t} %  kernel symbol for (anisotropic/isotropic) dilatation parameters. Was $\lambda$, but that is already used. 

\newcommand{\hubb}{\theta} % symbol for hubble rate (obv what it will be, but this aides searches for it)
\DeclareMathOperator\Vol{vol}

\newcommand\defn[1]{{\bfseries\slshape #1\/}}

\newcommand\Hg{H_{\mathrm{g}}}

\let\ring\mathring

\newcommand\bip{\gamma}
\newcommand\biA{\mspace{2mu}{}^\bip\mspace{-5mu}A}
\newcommand\bibA{{}^\bip\mspace{-5mu}\bar A}

\newcommand\biF{{}^\bip\mspace{-2mu}F}

\newcommand\bibF{{}^\bip\mspace{-2mu}\bar F}

\makeatletter
\def\@ApndToks#1#2{\edef\@act{\noexpand#1={\the#1#2}}\@act}
\def\@CopyToks#1#2{\edef\@act{\noexpand#2={\the#1}}\@act}

\newcommand{\dens}[2][1]{{
  \mathsurround=0pt  \everymath={\displaystyle}  \toks0={}
  \setbox0=\hbox{\lower 4.30554pt\hbox{$\mathchar"0365$}}  \dp0=0pt
  \count0=#1  \ifnum\count0 < 0  \multiply\count0 by -1  \fi
  \loop  \advance\count0 by -1  
    \@ApndToks{\toks0}{\copy0\crcr}
  \ifnum\count0 > 0
    \@ApndToks{\toks0}{\noalign{\kern -1pt\nointerlineskip}}
  \repeat
  \count0=#1  \skip3=0pt plus 3fil  \skip1=0pt plus 1fil
  \ifnum\count0 > 0
    \vbox{\ialign{\hskip\skip3##\hskip\skip1\crcr
      \the\toks0\noalign{\nointerlineskip}${#2}$\crcr}}
  \else
    \vtop{\ialign{\hskip\skip1##\hskip\skip3\crcr
      ${#2}$\crcr\noalign{\nointerlineskip\kern 2pt}\the\toks0}}
  \fi}\vphantom{#2}}

\makeatother

\newcommand{\pairpunct}[4][d]{% Arguments are: 1) The size of the punctuation mark(s) (default "d": Allow system to decide with \right and \left, "0": no size marker (just regular-sized punctuation), or "\size_command" ("\big" or whatever): use the size generated by that command), 2) The left punctuation mark ( (, [, etc), 3) The right puncuation mark, 4) The content to be enclosed in the punctuation marks.
\if#2.
\if#1d
\left. #4 \right#3
\else
\if#10
 #4 #3
\else
 #4 #1#3
\fi
\fi
\else
\if#3.
\if#1d
\left#2 #4 \right.
\else
\if#10
#2 #4 %\right.
\else
#1#2 #4 %\right.
\fi
\fi
\else
\if#1d
\left#2 #4 \right#3
\else
\if#10
#2 #4 #3
\else
#1#2 #4 #1#3
\fi
\fi
\fi
\fi
}
\newcommand{\pairparenth}[2][d]{\pairpunct[#1]{(}{)}{#2}}
\newcommand{\lambdacoord}{\lambda}	% The symbol for the coordinate set we called lambda
\newcommand{\deltop}{\hat{\Delta}}
\newcommand{\embedding}{{\iota}}

\begin{document}

\title{Quantum isotropy and the reduction of dynamics in \mbox{Bianchi I}}
\author{C Beetle${}^1$, J S Engle${}^1$, M E Hogan${}^{1,2,3,4}$, and P Mendon\c{c}a${}^1$}
\address{${}^1$Department of Physics, Florida Atlantic University, 777 Glades Road, Boca Raton, FL 33431, USA}
\address{${}^2$Department of Physics and Astronomy, Texas Tech University -- Costa Rica, Avenida Escaz\'{u}, Edificio AE205, San Rafael de Escaz\'{u}, San Jos\'{e}  10201, Costa Rica}
\address{${}^3$Department of Physics and Astronomy, Texas Tech University, Box 41051, Lubbock, TX 79409, USA}
\address{${}^4$Department of Mathematics and Statistics, Texas Tech University, 1108 Memorial Circle, Lubbock, TX 79409, USA}
\ead{cbeetle@fau.edu, jonathan.engle@fau.edu, matthew.hogan@ttu.edu, pmendon1@fau.edu}

\begin{abstract}
The authors previously introduced a diffeomorphism-invariant definition of a homogeneous and isotropic sector of loop quantum gravity, along with a program to embed loop quantum cosmology into it.  The present paper works out that program in detail for the simpler, but still physically non-trivial, case where the target of the embedding is the homogeneous, but not isotropic, Bianchi~I model.  The diffeomorphism-invariant conditions imposing homogeneity and isotropy in the full theory reduce to conditions imposing isotropy on an already homogeneous Bianchi~I spacetime.  The reduced conditions are invariant under the residual diffeomorphisms still allowed after gauge fixing the Bianchi~I model.  We show that there is a unique embedding of the quantum isotropic model into the homogeneous quantum Bianchi~I model that (a) is covariant with respect to the actions of such residual diffeomorphisms, and (b) intertwines both the (signed) volume operator and at least one directional Hubble rate.  That embedding also intertwines all other operators of interest in the respective loop quantum cosmological models, including their Hamiltonian constraints.  It thus establishes a precise equivalence between dynamics in the isotropic sector of the Bianchi I model and the quantized isotropic model, and not just their kinematics.  We also discuss the adjoint relationship between the embedding map defined here and a projection map previously defined by Ashtekar and Wilson-Ewing.  Finally, we highlight certain features that simplify this reduced embedding problem, but which may not have direct analogues in the embedding of homogeneous and isotropic loop quantum cosmology into full loop quantum gravity.
\end{abstract}

%\tableofcontents

\section{Introduction}

Quantum gravity is a domain of physics in which contact with observation remains a challenge, due to the extreme nature of the Planck scale where effects of the corresponding theory are expected to become relevant. That being said, due to cosmic expansion, the entire visible universe was once Planck sized. 
Indeed, cosmology has emerged as a 
promising domain in which to observe potential effects of quantum gravity
\cite{agullo2018, abs2017, agullo2015, aan2013}, 
and perhaps such effects have even already been observed \cite{aks2020, agjs2020}.

Loop quantum gravity (LQG) is a minimalist approach to a theory of quantum gravity guided foremost by Einstein's general principle of relativity, which in modern times is reformulated as diffeomorphism covariance, or background independence. Loop quantum cosmology (LQC) is a quantization of the homogeneous isotropic sector of gravity using the same techniques as loop quantum gravity.  To derive the effects of LQG on cosmology, the nearly exact homogeneity and isotropy of the early universe is exploited by using LQC for calculations. The relative simplicity of LQC allows for exact solutions to dynamics as well as the construction of a complete set of Dirac observable operators.

One can ask whether LQC,
a quantization of a symmetry reduced sector 
of gravity, accurately reflects the physics of
full loop quantum gravity.
When the choices made in the quantizations of a
field theory and its corresponding symmetry-reduced model are chosen to be appropriately compatible, symmetry reduction and quantization can indeed commute \cite{engle2006}.
In order to ask whether LQC reflects the appropriate sector of LQG, one must first specify what this sector is. It should be the quantum analogue of the homogeneous isotropic sector of classical gravity - that is, it should be the space of \textit{states} in LQG which are homogeneous and isotropic in some sense which is compatible with the diffeomorphism invariance of the theory. 
A proposal for such a sector has been defined in the prior work \cite{behm2017, behm2016} by finding diffeomorphism covariant phase space functions on the full gravity phase space whose vanishing is equivalent to the condition of homogeneity and isotropy with respect to \textit{some} maximal symmetry group on the spatial slice --- the \textit{symmetry conditions}.  These phase space functions are furthermore readily quantizable on the loop quantum gravity Hilbert space, so that the simultaneous kernel of the corresponding operators defines the desired sector in question. 
The second step is to find some \textit{embedding} of LQC states into the states of this sector. The work \cite{behm2017, behm2016} did this for a non-interacting toy example and sketched how to embed LQC into full LQG. 

The value of constructing an embedding of LQC into full LQG is not simply to both clarify the meaning of homogeneous isotropic in LQG as well as to 
understand how well LQC represents the physics of this sector. The value, more importantly, lies in its potential to associate each quantization choice in the full theory with a corresponding choice in the reduced theory. 
With such an association in hand, contact between LQC and observation can provide not only a test of LQG, but can also guide choices made in quantizing the full theory.
There are a number of programs which have been introduced to establish such an association \cite{dl2017, bodendorfer2016, bodendorfer2015, ac2016, ac2014, ydm2009, aps2006}.  
The advantages of the present program are that (1) it is compatible with the dynamics in the full theory, in the sense that diffeomorphism covariance is left intact without gauge fixing, and (2) it is compatible with the full space of states in LQG, in the sense that one does not need to restrict to states with support in a lattice. Since the so-called `$\overline{\mu}$-scheme' in LQC 
arises from requirements of diffeomorphism covariance 
\cite{ev2019, cs2008, aps2006},
it is reasonable to hope that the above two properties of the present strategy will enable a derivation of the $\overline{\mu}$-scheme from full LQG without inserting it by hand, in contrast to other approaches up until now.
Still, we expect there to be a relation between the approach followed here and at least the approaches of \cite{dl2017, ac2014}: 
The map from LQC states to (gauge-fixed, lattice-truncated) LQG states implicit in these latter approaches are based on coherent states, and the range of this implicit embedding is the span of all coherent states with homogeneous isotropic labels. This space is precisely the simultaneous kernel of quantum operators corresponding to the `holomorphic part' of the appropriate symmetry conditions \cite{engle2007, engle2006}, which are complex in a way exactly analogous to the complex symmetry conditions considered in the strategy of the present work. %\cite{behm2017, behm2016}.
%There is therefore hope that the programs in the works \cite{dl2017, ac2014}, which were successfully completed for at least a gauge-fixed lattice-truncated version of the full theory, will provide guidance in completing the present program. 

The goal of the present paper is to complete the program of \cite{behm2017, behm2016}, but in the simpler case of embedding LQC into \emph{Bianchi~I} LQC, in which homogeneity, but not isotropy, holds \textit{a priori}. The goal of doing this is to see how the program can be carried out to completion in this simpler, but still realistic case, thereby solidifying confidence in the program as well as providing an opportunity to gain intuition that will aid in applying it to embed into full LQG. 
The results turn out to be cleaner, more satisfactory, and more revealing than we had expected.

In the Bianchi~I model, the fully diffeomorphism-invariant condition imposing homogeneity and isotropy introduced in \cite{behm2017, behm2016} reduces to a \emph{residual} diffeomorphism-invariant condition imposing only \emph{isotropy}, which can be easily quantized in a manner similar to that suggested in \cite{behm2017, behm2016} for the full theory. We furthermore find that there exists a unique embedding from isotropic to \mbox{Bianchi I} LQC states that is covariant with respect to (canonical) residual diffeomorphisms, and also intertwines the operators in the two theories corresponding to the signed volume and a single directional Hubble rate. This uniquely determined embedding has image contained in the kernel of the quantum isotropy conditions. It furthermore intertwines the Hamiltonian constraints in the two theories, as well as all physically meaningful operators. Interestingly, it is precisely the adjoint of the projection from \mbox{Bianchi I} to isotropic LQC proposed by Ashtekar and Wilson-Ewing in \cite{aw2009}.

The rest of this paper is organized as follows. 
In section 2 we review the \mbox{Bianchi I} model as defined by Ashtekar and Wilson-Ewing in \cite{aw2009}. We then derive in section 3 the restriction, to the \mbox{Bianchi I} phase space, of the constraints proposed in \cite{behm2017} imposing diffeomorphism invariant homogeneity and isotropy. The Poisson brackets of these symmetry conditions among themselves are calculated with an eye toward quantum theory. The general quantization strategy presented in \cite{aw2009} is then used to provide symmetry constraint operators on the \mbox{Bianchi I} Hilbert space, 
whose simultaneous kernel defines the `quantum isotropic sector' of \mbox{Bianchi I}. Section 3 ends with a review of the isotropic model. In section 4, we derive the embedding of this model into the quantum isotropic sector of \mbox{Bianchi I}, and exhibit its properties. The successes of the results are sufficiently surprising that we devote section 5 to clarifying the classical origins of these successes.  Lastly we close with a discussion.

\section{Review of Bianchi I}
\label{s:BIrev}

\subsection{Classical Theory}
\label{BI:Classical}

The spacetime metric in the Bianchi I model has the form 
\begin{equation}\label{BI:met}
	\ed s^2 = - N^2(t)\, \ed t^2 + a_x^2(t)\, \ed x^2 + a_y^2(t)\, \ed y^2 + a_z^2(t)\, \ed z^2.
\end{equation}
The natural (co-)triad field on a homogeneous slice of constant $t$ is 
\begin{equation}\label{BI:tri}
	e_a^i(t) := a^{}_i(t)\, \ring e_a^i, 
	\qquad\text{where}\qquad
	\ring e_a^i := \ed x^i_a
\end{equation}
is the fiducial (co-)triad.  Note that there is no sum over the index $i$ in this definition of $e^i_a$.  We will write all such sums explicitly.  Meanwhile, the extrinsic curvature of a homogeneous slice is 
\begin{equation}\label{BI:exK}
	K_{ab}(t) = \frac{1}{2}\, \Lie_u\, q_{ab}(t)
		= \sum_i \frac{a_i(t)\, \dot a_i(t)}{N(t)}\, \ring e_a^i\, \ring e_b^i, 
\end{equation}
where $u = \smash[t]{\frac{1}{N(t)}\, \pdby{t}}$ is the future-directed, unit normal to the homogeneous slice.  We will omit any explicit $t$-dependence below.

Geometrically, the spatial coordinates $x^i$ in (\ref{BI:met}) can be defined as affine parameters along three mutually orthogonal congruences of parallel geodesics in the Euclidean spatial geometry of the Bianchi I model.  Moreover, the directions of those congruences are fixed in (\ref{BI:exK}) to coincide with the principal axes of the extrinsic curvature tensor $K_{ab}$.  Given appropriate Cauchy data for the Bianchi I model, consisting of a Euclidean spatial metric $q_{ab}$ and a homogeneous extrinsic curvature $K_{ab}$, the spatial coordinates so defined are unique up to (a) affine reparameterizations $\varphi_{(\vec m, \vec b)} : x^i \mapsto \tilde x^i := m_i\, x^i + b^i$ of each congruence, with each $m_i \ne 0$, and (b) permutations $\varphi_\pi : x^i \mapsto \tilde x^i := x^{\pi(i)}$ of the coordinate axes, with $\pi \in S_3$.  Any choice of such coordinates defines a canonical diffeomorphism from the spatial slice to $\Re^3$.  The present coordinate ambiguity therefore reflects the \defn{restricted diffeomorphism group} $\overline{\operatorname{Diff}} \approx (S_3 \ltimes \Re_{\!\times}^3 ) \ltimes \Re^3$ mapping $\Re^3$ to itself, \textit{i.e.}, the group of spatial diffeomorphisms that preserve the partial gauge-fixing conditions implicit in (\ref{BI:met}) and (\ref{BI:exK}).

The loop quantization of general relativity originates in the Ashtekar formulation of the classical theory.  The basic variables of that formulation are the densitized triad 
\begin{equation}\label{BI:Edef}
	E^a_i := \abs{\det e}\, e^a_i
		= \frac{\abs{a_x a_y a_z}}{a_i}\, \ring E^a_i, 
\end{equation}
and the Ashtekar connection with Barbero--Immirzi parameter $\bip$.  The latter is given by 
\begin{equation}\label{BI:Ared}
	\biA_a^i := \Gamma_a^i + \bip\, K_{ab}\, e^{bi} 
		= \frac{\bip\, \dot a_i}{N}\, \ring e_a^i, 
\end{equation}
where $\Gamma_a^i$ is the spin connection form for $e^a_i$, relative to a flat reference connection.  Spatial geometry is already flat in the Bianchi I model, so it is simplest to choose the reference connection to be the spin connection, whence $\Gamma_a^i = 0$.  The symplectic structure in Ashtekar gravity generally has the form 
\begin{equation}\label{BI:fullss}
	\Omega(\delta_1, \delta_2) 
		:= \frac{2}{\kappa\bip} \int_{\mathcal{V}} \delta_{[1}\! \biA_a^i\, \delta_{2]} E^a_i, 
\end{equation}
where $\kappa = 8 \pi G_{\text{Newton}}$.  The integral in (\ref{BI:fullss}) diverges when the field perturbations involved are homogeneous and the spatial slice $\mathcal{V}$ is not compact.  But, precisely due to that homogeneity, it then makes sense to restrict the integral to a compact fiducial cell $\mathcal{V}$, \textit{i.e.}, to a finite, rectangular volume with edges parallel to the coordinate axes \cite{aw2009}.  The symplectic structure then reduces to 
\begin{equation}\label{BI:redss}
% 	\Omega(\delta_1, \delta_2) = \frac{2}{\kappa\bip} \sum_i \delta_{[1} c^i\, \delta_{2]} p_i
% 	\qquad\text{or}\qquad 
	\Omega = \frac{1}{\kappa\bip} \sum_i \ed c^i \wedge \ed p_i, 
\end{equation}
where we have introduced the reduced phase space coordinates $(c^i, p_i)$ such that 
\begin{equation}\label{BI:cpdef}
	\biA_a^i =: c^i\, \frac{\ring e_a^i}{L_i} 
	\qquad\text{and}\qquad
	E^a_i =: \frac{L_i}{L_x L_y L_z}\, p_i\, \ring E^a_i.
\end{equation}
The coordinate lengths $L_i := \abs[\nml]{\Delta x^i}$ of the edges of the fiducial cell $\mathcal{V}$ enter these definitions to render the canonical coordinates independent of the initial choice of adapted coordinates $x^i$ in (\ref{BI:met}).  It will be convenient to exclude those points of the phase space corresponding to degenerate spatial geometries, \textit{i.e.}, having one or more of the $p_i$, and hence the volume, equal to zero. 
Such points are irrelevant in the usual limit used to make predictions, namely that of large fiducial cell volume, corresponding to removal of the infrared regulator
\cite{ev2019, as2011a}.
Our \defn{Bianchi I phase space} is therefore $\Gamma \cong \Re^3 \times \Re_{\!\times}^3$ topologically, where the second factor excludes the three coordinate planes in $\Re^3$ where at least one $p_i$ vanishes.

Now we consider the transformations of the Bianchi I phase space induced by the restricted spatial diffeomorphisms described above.  The phase-space transformation associated with a diffeomorphism $\varphi_{(\vec m, \vec b)}$ mapping each coordinate axis to itself follows immediately from the pull-backs 
\begin{equation}\label{BI:stran}
	\varphi^*_{(\vec m, \vec b)} \mathring e_a^i = m_i \mathring e_a^i
	\qquad\leadsto\qquad 
	\varphi_{(\vec m, \vec b)} : (c^i, p_i) \mapsto (\tilde c^i, \tilde p_i) 
		:= \biggl( m_i c^i, \frac{|m_x m_y m_z|}{m_i}\, p_i \biggr).
\end{equation}
The translation parameter $\vec b$ in $\varphi_{(\vec m, \vec b)}$ has no effect in phase space, as one would expect for a homogeneous model.  The situation is slightly more complicated for the diffeomorphisms $\varphi_\pi : \Re^3 \to \Re^3$ that interchange the coordinate axes because  $\varphi_\pi^* \mathring e^i = \mathring e^{\pi(i)}$, which generally differs from $\mathring e^i$.  Thus, whereas the definition (\ref{BI:cpdef}) of the coordinates $(c^i, p_i)$ presumes that $A^x \propto \ed x$, $A^y \propto \ed y$, and so forth, the pullback $\varphi_\pi^* A_a^i$ no longer necessarily satisfies this parallelism condition.  This difficulty is easy to fix, however, by incorporating an appropriate, internal gauge rotation $R \in SO(3)$ such that 
\begin{align*}
	\varphi_{(\pi, R)} : \biA_a^i \mapsto \mspace{2mu}{}^\bip\mspace{-5mu}\tilde{A}^i_a 
		:= \sum_j R^i{}_j\, \varphi^*_\pi \biA_a^j 
		= \sum_j R^i{}_j\, c^j\, \frac{\mathring e^{\pi(j)}_a}{L_j} 
\end{align*}
is again proportional to $\mathring e^i_a$, and similarly for the physical (co-)triad $e^i_a$.  The rotation here must be chosen such that $R^i{}_j = 0$ unless $\pi(j) = i$.  The set of rotations mapping the coordinate axes into one another like this form the (chiral) octahedral group $O \subset SO(3)$, \textit{i.e.}, the subgroup of rotations preserving the unit cube.  For any fixed $\pi \in S_3$, there are exactly four rotations satisfying the above condition, differing from one another by half-rotations about one of the coordinate axes.  Choosing any one of them leads to 
\begin{align*}
	\varphi_{(\pi, R)} : (c^i, p_i) \mapsto (\tilde c^i, \tilde p_i) 
		:= \biggl( m_i\, c^{\pi^{-1}(i)}, \frac{1}{m_i}\, p_{\pi^{-1}(i)} \biggr) 
	\qquad\text{with}\qquad 
	m_i := R^i{}_{\pi^{-1}(i)}\, \frac{L_i}{L_{\pi^{-1}(i)}}.
\end{align*}
(Note that $R^i{}_{\pi^{-i}(i)} = \pm 1$ and $|m_x m_y m_z| = 1$ by definition.)  Composing with an appropriate scaling transformation from (\ref{BI:stran}) thus leads to a transformation that simply permutes the $(c^i, p_i)$ coordinates in pairs.

The \defn{residual automorphism group} $\operatorname{Aut}_R$ is the set of \emph{distinct} phase-space transformations induced by the restricted diffeomorphisms described above.  In detail, $\operatorname{Aut}_R \cong [\overline{\operatorname{Diff}} \times SO(3)]_\shortparallel / K_\Gamma$ is isomorphic to the group of restricted diffeomorphisms, extended to include (homogeneous) internal gauge rotations, then restricted to preserve the parallelism of (\ref{BI:cpdef}), and finally quotiented by the (normal) subgroup $K_\Gamma$ of such transformations that act as the identity in phase space.  The resulting group is naturally a semi-direct product $\operatorname{Aut}_R = \bigl( \operatorname{Dil}_R \times \operatorname{Par}_R) \rtimes \operatorname{Rot}_R$ of three distinct factors, consisting of 
\begin{enumerate}
\item \defn{anisotropic dilatations} $\varphi_{\vec t} \in \operatorname{Dil}_R \cong \Re_{\!+}^3$, labeled by $\vec t \in \Re^3$ and having the form 
\begin{equation}\label{BI:anis}
	\varphi_{\vec t}(c^i, p_i) := \big( \ee^{- t_i}\, c^i, \ee^{t_i - T}\, p_i \bigr) 
	\qquad\text{with}\qquad 
	T := \textstyle\sum_i t_i; 
\end{equation}
\item \defn{partial reflections} $\varphi_{\vec\zeta} \in \operatorname{Par}_R \cong S_2^3$, labeled by $\vec\zeta \in \{\pm 1\}^3$ and having the form 
\begin{equation}\label{BI:pref}
	\varphi_{\vec\zeta}(c^i, p_i) := \bigl( \zeta_i\, c^i, \zeta_i\, p_i \bigr); 
\end{equation}
\item and \defn{residual rotations} $\varphi_\pi \in \operatorname{Rot}_R \cong S_3$, labeled by $\pi \in S_3$ and having the form 
\begin{equation}\label{BI:rrot}
	\varphi_\pi(c^i, p_i) := \bigl( c^{\pi(i)}, p_{\pi(i)} \bigr).
\end{equation}
\end{enumerate}
Note that the partial reflections and residual rotations together define a natural action of the (achiral) octahedral group $\operatorname{Par}_R \rtimes \operatorname{Rot}_R \cong O_h \subset O(3)$, which is the full isometry group of the unit cube, including reflections.  The residual automorphism group has a non-trivial center $Z(\operatorname{Aut}_R)$, consisting of 
\begin{enumerate}
\item[(i$_S$)] \defn{isotropic dilatations} $\varphi_T \in \operatorname{Dil}_S \cong \Re_{\!+}$, labeled by $T \in \Re$ and having the form 
\begin{equation}\label{BI:idil}
	\varphi_T(c^i, p_i) := \bigl( \ee^{-T/3}\, c^i, \ee^{-2T / 3}\, p_i \bigr); 
\end{equation}
\item[(ii$_S$)] and \defn{isotropic reflections} $\varphi_Z \in \operatorname{Par}_S \cong S_2$, labeled by $Z \in \{\pm 1\}$ and having the form 
\begin{equation}\label{BI:ipar}
	\varphi_Z(c^i, p_i) := \bigl( Z\, c^i, Z\, p_i \bigr).
\end{equation}
\end{enumerate}
We refer to $Z(\operatorname{Aut}_R) \triangleleft \operatorname{Aut}_R$ as the \defn{isotropic automorphism group} not only because its elements ``act isotropically'' in the Bianchi I phase space, but also because it is naturally isomorphic to the group of residual automorphisms analogous to $\operatorname{Aut}_R$ for the fully reduced, isotropic model to be discussed in the next section.  The quotient group $\operatorname{Aut}_R / Z(\operatorname{Aut}_R)$ plays a pivotal role in relating the Bianchi I model to its isotropic reduction.  This quotient can be identified with the (normal) subgroup $\operatorname{Aut}_R^0 \triangleleft \operatorname{Aut}_R$ having $T = 0$ in (\ref{BI:anis}) and $\zeta_x \zeta _y \zeta_z = 1$ in (\ref{BI:pref}).  We refer to this as the \defn{proper residual automorphism group} because its elements preserve both the symplectic structure (\ref{BI:redss}) and the orientation of the physical triad (\ref{BI:tri}).

Turning now to the dynamics of the Bianchi I model, recall that the gravitational part of the classical Hamiltonian constraint involves the curvature of the homogeneous Ashtekar connection (\ref{BI:Ared}): 
\begin{equation}\label{BI:Fred}
	\biF_{ab}{}^i 
		:= \ed\biA_{ab}^i + \sum_{jk} \epsilon^i{}_{jk}\, \biA_a^j\, \biA_b^k
		= \frac{L_i}{L_x L_y L_z}\, \frac{c^x c^y c^z}{c^i} 
			\sum_{jk} \epsilon^i{}_{jk}\, \ring e_a^j\, \ring e_b^k.
\end{equation}
The coordinate scales $L_i$ enter because the result is expressed in terms of the fiducial triad.  One can relate it instead to the physical triad by solving for the original scale factors $a_i$: 
\begin{equation}\label{BI:aofp}
	p_i = \frac{L_x L_y L_z}{L_i}\, \frac{\abs{a_x a_y a_z}}{a_i}
	\qquad\leadsto\qquad
	a_i = \frac{1}{L_i}\, \frac{\Vol(\vec p)}{p_i} 
	\qquad\text{with}\qquad 
	\Vol(\vec p) := \abs{p_x p_y p_z}^{1/2}.
\end{equation}
Geometrically, $\Vol(\vec p)$ is the proper volume of the fiducial cell.  Substituting into (\ref{BI:Fred}) then gives 
\begin{equation}\label{BI:FSig}
	\biF_{ab}{}^i = \frac{c^x c^y c^z}{c^i}\, \frac{\sgn (p_x p_y p_z)}{p_i}\, \Sigma_{ab}{}^i
	\qquad\text{with}\qquad
	\Sigma_{ab}{}^i := \sum_{jk} \epsilon^i{}_{jk}\, e_a^j\, e_b^k.
\end{equation}
This result can be expressed compactly in terms of the \defn{directional Hubble rates} 
\begin{equation}\label{BI:dHub}
	\theta_i :=  \frac{\Lie_u a_i}{a_i} = \frac{\dot a_i}{N\, a_i} 
		= \frac{c^i\, p_i}{\bip \Vol(\vec p)} 
	\qquad\leadsto\qquad 
	\biA_a^i = \bip\, K_a^i = \bip\, \theta_i\, e_a^i.
\end{equation}
Each Hubble rate is invariant under anisotropic dilatations (\ref{BI:anis}) and partial reflections (\ref{BI:pref}), and they permute covariantly under the residual rotations of (\ref{BI:rrot}).  Meanwhile, the curvature of (\ref{BI:FSig}) is given by 
\begin{equation}\label{BI:FHub}
	\biF_{ab}{}^i = \bip^2\, \frac{\theta_x \theta_y \theta_z}{\theta_i}\, \Sigma_{ab}{}^i
		= \bip^2 \sum_{jk} \epsilon^i{}_{jk}\, \theta_j\, \theta_k\, e^j_a\, e^k_b.
\end{equation}
This yields a compact expression for (the gravitational part of) the Hamiltonian constraint: 
\begin{align}\label{BI:Hred}
	\Hg[N]
		:={}&{} \frac{1}{2 \kappa} \int_{\mathcal{V}}\, \sum_{ij} \Biggl( 
			\frac{N\, E^a_i\, E^b_j}{\abs{\det E}^{1/2}} 
			\biggl( \sum_k \epsilon^{ij}{}_k\, \biF_{ab}{}^k - 2 (1 + \bip^2)\, K_{[a}{}^i\, K_{b]}{}^j \biggr)
			\Biggr)
			\notag\\[1ex]
		={}&{} - \frac{\sgn(\det e)}{\kappa \bip^2} \int_{\mathcal{V}} N \sum_k \biF^k \wedge e_k 
			\notag\\[1ex] 
		={}&{} - \frac{\Vol[N](\vec p)}{\kappa}\, \sum_{i < j} \theta_i\, \theta_j.
\end{align}
The two terms in the integrand on the first line are proportional to one another in the homogeneous case because $\ed \biA_{ab}{}^i = 0$.  Meanwhile, $\Vol[N](\vec p)$ denotes the lapse $N$ integrated over the fiducial cell using the proper volume element determined by $\vec p$.  Note that $\Hg[N]$ is invariant under the same subgroup of residual automorphisms that preserve the symplectic structure (\ref{BI:redss}), and that its Hamiltonian flow preserves the submanifold of homogeneous states in the phase space of the full theory if and only if $N$ is homogeneous.

\subsection{Regularization Strategy}

The curvature (\ref{BI:Fred}) has no operator analogue in the Ashtekar--Lewandowski quantization of gravity.  One therefore introduces a \textit{regularized} curvature at the classical level for the Bianchi I model \cite{aw2009}.  The regularized curvature is constructed from holonomies along finite curves, which do have operator analogues in the full theory.  This classical regularization should therefore be viewed as a part of the quantization process.

It is natural in Bianchi I to consider the holonomy around a rectangular loop with edges parallel to two of the three principal curvature axes.  Working in the fundamental representation of SU(2), the holonomy of the connection (\ref{BI:Ared}) along a line segment of coordinate length $\ell$ parallel to the $x^i$-axis is given by 
\begin{align*}
	h_i(\ell) = 1 \cos \frac{c^i \ell}{2 L_i} + 2 \tau_i \sin \frac{c^i \ell}{2 L_i}, 
\end{align*}
where $1$ denotes the $2 \times 2$ identity matrix, and $\tau_i$ are the anti-Hermitian generators of the fundamental representation of SU(2), related to the Pauli matrices by $\sigma_i =  2 \ii \tau_i$.
% These are related to the Pauli matrices by $\sigma_i =  2 \ii \tau_i$, so 
% %
% \begin{align*}
% 	\tau_i\, \tau_j = \frac{1}{2} \sum_k \epsilon_{ij}{}^k\, \tau_k - \frac{1}{4}\, \delta_{ij}\, 1.
% \end{align*}
% %
It follows that the holonomy around a closed, rectangular plaquette with edges of coordinate lengths $\ell_i$ and $\ell_j$ parallel to the $x^i$- and $x^j$-axes, respectively, is 
\begin{align*}
	h_{ij}(\vec\ell) 
		:={}&{} h_j(-\ell_j)\, h_i(-\ell_i)\, h_j(\ell_j)\, h_i(\ell_i) 
			\notag\\[1ex]
		={}&{} \biggl( 1 - 2 \sin^2 \frac{c^i \ell_i}{2 L_i}	\sin^2 \frac{c^j \ell_j}{2 L_j} \biggr) 1 
			+ 2 \sin \frac{c^i \ell_i}{L_i} \sin^2 \frac{c^j \ell_j}{2 L_j}\, \tau_i 
			\notag\\&\hspace{3em}
			- 2 \sin^2 \frac{c^i \ell_i}{2 L_i} \sin \frac{c^j \ell_j}{L_j}\, \tau_j 
			- \sin \frac{c^i \ell_i}{L_i} \sin \frac{c^j \ell_j}{L_j}\, \comm{\tau_i}{\tau_j}, 
\end{align*}
where $\vec\ell := (\ell_x, \ell_y, \ell_z)$.  The last term here is quadratic as $\vec\ell \to \vec 0$, and proportional to the curvature $\biF_{ab}{}^k$ from (\ref{BI:Fred}) in that limit.  More precisely, we have 
\begin{equation}\label{BI:Flim}
	\lim_{\vec\ell \to \vec 0} \sum_{ij} \Tr \Bigl[ h_{ij}(\vec\ell)\, \tau^k \Bigr]\, \frac{\mathring e_a^i}{\ell_i}\, \frac{\mathring e_b^j}{\ell_j} \to \frac{1}{2}\, \biF_{ab}{}^k
\end{equation}
It is tempting to define the regularized curvature simply by removing the limit.  But the continuum curvature $\biF_{ab}{}^k$ from (\ref{BI:Fred}) has the property that its pull-back to the plane orthogonal to $e_k^a$ is proportional to $\tau^k$, whereas the expression under the limit in (\ref{BI:Flim}) does not.  Happily, the cubic terms that create this difficulty cancel one another if instead we set 
\begin{equation}\label{BI:Freg}
	\biF_{ab}{}^k(\vec\ell) 
		:= \sum_{ij} \Tr \Bigl[ \bigl( h_{ij}(\vec\ell) + h_{ij}(-\vec\ell) \bigr)\, \tau^k \Bigr]\, \frac{\mathring e_a^i}{\ell_i}\, \frac{\mathring e_b^j}{\ell_j} 
		= \sum_{ij} \sin \frac{c^i \ell_i}{L_i} \sin \frac{c^j \ell_j}{L_j}\, \epsilon_{ij}{}^k\, \frac{\mathring e_a^i}{\ell_i}\, \frac{\mathring e_b^j}{\ell_j}.
\end{equation}
This regularized curvature is amenable to quantization, though again it depends on the unphysical, coordinate lengths of the plaquette edges.

The Bianchi I model is spatially homogeneous, so the proper length $s_i$ of any line segment along a symmetry axis is proportional to its coordinate length $\ell_i$.  Accordingly, we may set 
\begin{align*}
	\ell_i\, \ring e^a_i = s_i\, e^a_i 
	\qquad\leadsto\qquad
	\frac{\ell_i}{L_i} = \frac{p_i\, s_i}{\Vol(\vec p)} 
\end{align*}
to define a vector $\vec\ell$ of coordinate edge lengths corresponding to a given vector $\vec s$ of proper edge lengths.  The holonomy along an edge of fixed proper length is denoted 
%
% N.B. in the literature the holonomy is usually defined with 
% absolute value around $p_i$.  This has the advantages of being 
% covariant under certain reflections, the resulting operators 
% have consistent composition properties at least along
% the same directions.
% I want to change this before submitting to CGQ - that means 
% changing in the above equation defining $\ell_i$,
% below, and propagating it further.  I suspect we won't need to 
% propagate far, as one will have the final same Hamiltonian 
% constraint.
%
\begin{equation}\label{BI:bhol}
	\bar h_i(s) = 1 \cos \frac{c^i\, p_i\, s}{2 \Vol(\vec p)} 
		+ 2 \tau_i \sin \frac{c^i\, p_i\, s}{2 \Vol(\vec p)}.
\end{equation}
Recasting the regularized curvature (\ref{BI:Freg}) in terms of proper lengths gives 
\begin{equation}\label{BI:Fhol}
	\bibF_{ab}{}^k(\vec s) 
		:= \sum_{ij}{} \Tr \Bigl[ \bigl( \bar h_{ij}(\vec s) + \bar h_{ij}(- \vec s) \bigr)\, \tau^k \Bigr]\, 
			\frac{e_a^i}{s_i}\, \frac{e_b^j}{s_j}
		= \sum_{ij}{} \sin \frac{c^i p_i s_i}{\Vol(\vec p)} \sin \frac{c^j p_j s_j}{\Vol(\vec p)}\, 
			\epsilon_{ij}{}^k\, \frac{e_a^i}{s_i}\, \frac{e_b^j}{s_j}.
\end{equation}
This expression coincides, for appropriate choices of the $s_i$ to be described in the next subsection, with the regularized curvature obtained in \cite{aw2009, wilsonewingthesis}, though the latter is written in terms of its components relative to the \textit{fiducial} triad basis.  Comparing with (\ref{BI:FHub}), the present expression suggests defining a regularized version of the directional Hubble rate by setting 
\begin{equation}\label{BI:rhub}
	\bar\theta_i(s) := \frac{1}{\bip s} \sin \frac{c^i p_i s}{\Vol(\vec p)}.
\end{equation}
Note that, like the ordinary directional Hubble rates (\ref{BI:dHub}), these expressions are invariant under anisotropic dilatations (\ref{BI:anis}) and partial reflections (\ref{BI:pref}), and permute covariantly under residual rotations (\ref{BI:rrot}).

The standard construction of the Hamiltonian constraint in full loop quantum gravity, due to Thiemann \cite{thiemann1996}, begins by observing that 
\begin{align*}
	e^k_c(x)
		= \frac{2}{\kappa \bip}\, \frac{\pbrack[\nml]{A_c^k(x)}{\Vol[N]}}{N(x)}.
\end{align*}
The connection in the Poisson bracket can be regularized in terms of holonomies by noting that, for an arbitrary curve $\xi(t)$, one has 
\begin{align*}
	\dot\xi^c(0)\, \pbrack[\big]{\biA_c^k \bigl( \xi(0) \bigr)}{\Vol[N]}
		= - 2 \Tr \biggl[ \tau^k\, \fdby{t} \Bigl( h_\xi^{-1}(t)\, \pbrack{h_\xi^{}(t)}{\Vol[N]} \Bigr)_{t = 0} \biggr].
\end{align*}
We take $\xi(t)$ to run along one of the principal axes in the Bianchi I case, and parameterize the curve by proper length.  Approximating the derivative with a finite difference then leads to 
\begin{align*}
	\pbrack[\big]{\bibA_c^k}{\Vol[N](\vec p)}(\vec s) := - 2\, \sum_l\, \frac{e^l_c}{s_l} 
		\Tr \Bigl[ \tau^k\, \bar h_l(-\vec s)\, \pbrack{\bar h_l(\vec s)}{\Vol[N](\vec p)} \Bigr].
\end{align*}
Substituting this result and the regularized curvature (\ref{BI:Fhol}) into the Hamiltonian constraint from the second line of (\ref{BI:Hred}) gives the regularized Hamiltonian constraint 
\begin{equation}\label{BI:Hhol}
	\Hg[N](\vec s) = - \frac{\Vol[N](\vec p)}{\kappa^2 \bip^3}\, \sum_{ijk}\, \frac{\epsilon_{ijk}}{s_i s_j s_k}
		\Tr \Bigl[ \Bigl( \bar h_{ij}(\vec s) + \bar h_{ij}(- \vec s) \Bigr)\, 
		\bar h_k(-\vec s)\, \pbrack{\bar h_k(\vec s)}{\Vol(\vec p)} \Bigr], 
\end{equation}
where we have used the identity 
\begin{align*}
	\sum_k \Tr(X \tau^k) \Tr(Y \tau_k) = \frac{\Tr(X) \Tr(Y) - 2 \Tr(XY)}{4}.
\end{align*}
To compare this result to the expression from \cite{aw2009, wilsonewingthesis}, one can explicitly calculate the Poisson bracket 
\begin{align*}
	\pbrack{\bar h_k(s)}{\Vol(\vec p)} =\frac{\kappa\bip s}{2}\, \tau_k\, \bar h_k(s).
\end{align*}
Since $\tau_k$ commutes with $\bar h_k(s)$, the trace in (\ref{BI:Hhol}) reduces to the same one from (\ref{BI:Fhol}), yielding 
\begin{equation}\label{BI:Hreg}
	\Hg[N](\vec s) = - \frac{{\Vol[N](\vec p)}}{\kappa}\, \sum_{i < j}\, 
			\biggl( \frac{1}{\bip s_i} \sin \frac{c^i p_i s_i}{\Vol(\vec p)} \biggr)
			\biggl( \frac{1}{\bip s_j} \sin \frac{c^j p_j s_j}{\Vol(\vec p)} \biggr).
\end{equation}
Thus, the regularized Hamiltonian constraint has exactly the form of the last expression from (\ref{BI:Hred}), but with the directional Hubble rates $\theta_i$ replaced by their regularized analogues $\bar\theta_i(\vec s)$ from (\ref{BI:rhub}).  Once again this expression coincides, for appropriate choices of the $s_i$, with the Hamiltonian constraint from \cite{aw2009, wilsonewingthesis}.  (Note that the latter is presented only in the harmonic time gauge, where the lapse $N = N_0 \Vol(\vec p)$ is spatially constant, but state-dependent.)

\subsection{Quantum Theory}
\label{BI:Quantum}

Quantizing the regularized expressions from the previous subsection entails promoting holonomies to quantum operators, and specifying operator orderings where ambiguities arise.  To quantize the holonomies (\ref{BI:bhol}), Ashtekar and Wilson-Ewing define the complex exponentials 
\begin{equation}\label{BI:Ddef}
	\Delta_i(s) := \exp - \frac{\ii c^i \abs{p_i} s}{2 \Vol(\vec p)}
\end{equation}
at the classical level.  Note the absolute value $|p_i|$ in the numerator here, which renders $\Delta_i(s)$ invariant under anisotropic dilatations (\ref{BI:anis}), but \emph{not} under partial reflections (\ref{BI:pref}).  Instead, we have $\varphi_{\vec\zeta}^* \Delta_i(s) = \Delta_i(\zeta_i s)$.  This seemingly undesirable asymmetry is critical for quantization, as we now show.

One \textit{motivates} the quantization of the functions (\ref{BI:Ddef}) by recalling that the classical coordinate $c^i$ becomes a derivative operator in a Schr\"odinger representation based on $\vec p$: 
\begin{equation}\label{BI:ordD}
	c^i \mapsto \ii\hbar\kappa\bip\, \pdby{p_i}
	\qquad\leadsto\qquad
	- \frac{\ii \abs{p_i} c^i s}{2 \Vol(\vec p)} \mapsto \frac{\hbar\kappa\bip\, \abs{p_i}^{1/2} s}{2\, \abs{p_x p_y p_z}^{1/2}} \cdot \abs{p_i}^{1/2}\, \pdby{p_i}.
\end{equation}
The first factor on the right is independent of $p_i$, and thus can be treated as a constant along each orbit of this vector field on $\Re^3$.  The second factor can be affinely parameterized such that 
\begin{equation}\label{BI:affp}
	\abs{p_i}^{1/2}\, \pdby{p_i} = \frac{\abs{p_0}^{1/2}}{2 p_0}\, \pdby{\lambda_i}
	\qquad\text{with}\qquad
	p_i =: p_0 P(\lambda_i) := p_0 \sgn(\lambda_i)\, \lambda_i^2, 
\end{equation}
where $p_0$ is an arbitrary constant with units of area to be fixed below. The resulting affine parameter $\lambda_i$ is dimensionless and increases (as long as $p_0 > 0$) monotonically with $p_i$.  Thus, a Schr\"odinger representation based on $\vec p$ is closely related to a Schr\"odinger representation based on $\vec\lambda$, though the two have different natural inner products since $\ed^3 p = \abs{p_0}^{3/2} \Vol(\vec p)\, \ed^3 \lambda$.  More importantly, however, the vector field in question is a constant multiple of $\pdby{\lambda_i}$ on each of its integral curves, and therefore generates a rigid translation in $\lambda_i$.  It follows that the natural (Schr\"odinger) quantization of $\Delta_i(s)$ is such that 
\begin{equation}\label{BI:braDpre}
	\bra[\big]{\vec\lambda} \hat\Delta_i(s)
		:= \bra{\vec \lambda + \frac{\hbar\kappa\bip s_0}{4 p_0 \abs{p_0}{}^{1/2}}\, \frac{\abs{\lambda_i}}{\abs{\lambda_x \lambda_y \lambda_z}}\, \frac{s}{s_0}\, \vec e_i}, 
\end{equation}
where $\vec e_i$ is the canonical basis vector in $\Re^3$ and $s_0$ is a length scale to be fixed below.  As usual, the dual basis vectors $\bra{\vec\lambda}$ here map a state $\ket{\psi}$ to its value $\psi(\vec\lambda)$ at a particular point $\vec\lambda \in \Re^3$.  Note that, if we were to replace $|p_i|$ with $p_i$ on the right side of (\ref{BI:Ddef}), then the flow of this vector field would reverse in the half-space $\lambda_i < 0$, and therefore would not be \emph{globally} integrable \cite{wilsonewingthesis}.  (For $s < 0$, for example, the flow would converge on the plane $\lambda_i = 0$ from both sides in finite affine parameter ``time,'' and one cannot continue to integrate through that plane.)

Although the Schr\"odinger representation based on $\vec p$ motivates the quantization (\ref{BI:braDpre}) of $\Delta_i(s)$, the resulting operator needs to act in the ``polymer'' Hilbert space \cite{aw2009} of loop quantum cosmology.  The inner product on this space is the \textit{sum} 
\begin{equation}\label{BI:polyip}
	\iprod{\phi}{\psi} 
		= \sum_{\vec p} \bar\phi(\vec p)\, \psi(\vec p) 
		= \sum_{\vec\lambda} \bar\phi(\vec\lambda)\, \psi(\vec\lambda).
\end{equation}
The distinction is important.  The ordering $\abs{p_i}^{1/2}\, c^i \mapsto \abs[\nml]{\hat p_i}^{1/2}\, \hat c^i$ chosen in (\ref{BI:ordD}) is the unique one that leaves a \textit{constant} wave function $\psi(\vec p) = \psi_0$ invariant under the action of the resulting translation operator.  But this ordering is not Hermitian in the Schr\"odinger representation based on $\vec p$, and its exponential $\hat\Delta_i(s)$ is not unitary: this is because the Lebesgue measure $\ed^3 p$ is not invariant under a rigid translation in $\vec\lambda$.  The polymer representations based on $\vec p$ and $\vec\lambda$ are the same, however, so (\ref{BI:braDpre}) \textit{is} unitary in loop quantum cosmology.

Typically one would take the limit $\vec s \to \vec 0$ after quantization to remove the regulator in (\ref{BI:braDpre}), but that limit does not exist in loop quantum cosmology.  Instead, one fixes a  certain \textit{finite} value of $\vec s$ to define \textit{the} curvature and Hamiltonian constraint operators by setting 
\begin{align}
	s_0^2 := \hbar\kappa \abs{\bip} \sqrt{j_0 (j_0 + 1)}, 
\end{align}
where it is natural to take $j_0 = \frac{1}{2}$ so that $s_0^2$ is the minimal quantum of area in full loop quantum gravity.  This fixes the length scale introduced in (\ref{BI:braDpre}).  Then one chooses the area scale $p_0$ from (\ref{BI:affp}) such that the ratio of dimensional factors in (\ref{BI:braDpre}) is one half: 
\begin{equation}\label{BI:p0val}
	p_0 = \frac{\hbar\kappa\bip}{2} \sqrt[6]{4 j_0 (j_0 + 1)}.
\end{equation}
With these choices, the basic operators of loop quantum cosmology act according to 
\begin{equation}\label{BI:braDs}
	\bra[\big]{\vec\lambda} \hat\Delta_i(s) := \bra{\vec\lambda + \frac{\abs{\lambda_i}}{\abs{\lambda_x \lambda_y \lambda_z}}\, \frac{s}{2 s_0}\, \vec{e}_i} 
	\qquad\text{and}\qquad 
	\bra[\big]{\vec\lambda} \hat p_i :=  p_0\, P(\lambda_i)\, \bra[\big]{\vec\lambda}, 
\end{equation}
where again $P(\lambda_i) := \sgn(\lambda_i)\, \lambda_i^2$.  The regularization scheme for the Hamiltonian constraint in \cite{aw2009, wilsonewingthesis}, which we extend here to other operators that are needed to enforce the quantum symmetry conditions, simply sets all $s_i = s_0$.  Accordingly, we introduce the shorthands 
\begin{equation}\label{BI:braD}
	\hat\Delta_i := \hat\Delta_i(s_0)
	\qquad\leadsto\qquad 
	\bra[\big]{\vec\lambda} \hat\Delta_i := \bra{\vec\lambda + \frac{\abs{\lambda_i}}{2 \abs{\Lambda}}\, \vec e_i} 
	\qquad\text{with}\qquad 
	\Lambda := \lambda_x \lambda_y \lambda_z.
\end{equation}
For purposes of comparison, the basic holonomy operators $\hat E_i^\pm$ from \cite{wilsonewingthesis}\relax
\footnote{The same notation, $\hat E_i^-$, is used in \cite{aw2009} to denote a slightly different operator, which omits the absolute value in the denominator from (\ref{BI:braD}).  That operator is not unitary, however.  See \cite{wilsonewingthesis} for a corrected expression.}
correspond to $\hat\Delta_i^{\pm 2}$ in the notation we use here.  Note that the effect of $\hat\Delta_i^{\pm 2}$ is to shift only the $\lambda_i$ component of the argument of the given wave function $\psi(\vec\lambda) := \braket{\vec\lambda}{\psi}$ such that $\Lambda \mapsto \Lambda^\pm := \Lambda \pm \sgn(\Lambda\lambda_i)$.

The standard approach in loop quantum cosmology is to reduce the regularized Hamiltonian constraint (\ref{BI:Hhol}) to the scalar form (\ref{BI:Hreg}) \emph{prior} to quantization.  To do this, write (\ref{BI:Hreg}) in the form 
\begin{equation}\label{BI:Hredb}
	H_{\mathrm{g}}[N](\vec s) = - \frac{\Vol[N](\vec p)}{\kappa} \sum_{i < j} 
		\biggl( \frac{\sgn p_i}{\bip s_i}\, \frac{\Delta_i(2 s_i) - \Delta_i(- 2 s_i)}{2 \ii} \biggr)
		\biggl( \frac{\sgn p_j}{\bip s_j}\, \frac{\Delta_j(2 s_j) - \Delta_j(- 2 s_j)}{2 \ii} \biggr).
\end{equation}
The operator analogues of the various factors in this expression do not commute, and one must choose how to order them in defining the operator analogue of the Hamiltonian constraint.  Ashtekar and Wilson-Ewing \cite{aw2009} first choose a Weyl ordering for each factor in braces, the directional Hubble rates from (\ref{BI:rhub}), defining the corresponding operators\relax 
\footnote{The present notation suggests a simpler ordering, namely $\hat\theta_i := \bigl( \hat\Delta_i(-s) \sgn\hat\lambda_i\, \hat\Delta_i(-s)- \hat\Delta_i(s) \sgn\hat\lambda_i\, \hat\Delta_i(s) \bigr) / 2 \ii \bip s$. However, 
the choice of ordering does not matter in the limit of large fiducial cell volume (removal of the infrared regulator) usually used to extract physical predictions \cite{ev2019, as2011a}, whence we retain the conventional ordering of \cite{aw2009}.}
\begin{align}
	\hat\theta_i(s_i)
		:={}&{} \frac{\sgn\hat\lambda_i\, (\hat\Delta_i(-2s_i) - \hat\Delta_i(2s_i)) 
			+ (\hat\Delta_i(-2s_i) - \hat\Delta_i(2s_i)) \sgn\hat\lambda_i}
				{4 \ii \bip s_i}
			\notag\\[1ex]
		={}&{} \Bigl(
		 \hat\Delta_i(-2s_i) \,
			\Theta \big( \abs[\nml]{\hat\Lambda} + (s_i/s_0)\sgn\hat\lambda_i\bigr)
			- \hat\Delta_i(2s_i) \,
			\Theta \bigl( \abs[\nml]{\hat\Lambda} - (s_i/s_0)\sgn\hat\lambda_i\bigr)\Bigr)\,\frac{\sgn\hat\lambda_i}{2 \ii \bip s_i}
     \label{BI:dHops}\\[1ex]
      ={}&{}\frac{\sgn\hat\lambda_i}{2 \ii \bip s_i}\, \Bigl(
			\Theta \big( \abs[\nml]{\hat\Lambda} - (s_i/s_0)\sgn\hat\lambda_i\bigr)\, \hat\Delta_i(-2s_i) 	
			- \Theta \bigl( \abs[\nml]{\hat\Lambda} + (s_i/s_0)\sgn\hat\lambda_i\bigr)\, \hat\Delta_i(2s_i)\Bigr),
     \notag
\end{align}
where $\Theta(t)$ denotes the Heaviside step function. 
Setting $s_i:= s_0$ for all $i=x,y,z$, these become
\begin{align}
\label{BI:dHop}
\hat\theta_i= \frac{\sgn\hat\lambda_i}{2 \ii \bip s_0}\, \Bigl(
			\Theta \big( \abs[\nml]{\hat\Lambda} - \sgn\hat\lambda_i \bigr)\, \hat\Delta_i^{-2} 	
			- \Theta \bigl( \abs[\nml]{\hat\Lambda} + \sgn\hat\lambda_i \bigr)\, \hat\Delta_i^{+2}\Bigr). 
\end{align}
Ashtekar and Wilson-Ewing then distribute the volume factor from outside the sum in (\ref{BI:Hred},\ref{BI:Hredb}) symmetrically and again choose a Weyl ordering for the product\relax 
\footnote{Even classically one has $	\pbrack{\theta_i}{\theta_j} = \kappa\, (\theta_i - \theta_j) / 2 \Vol(\vec p)$, so some ordering prescription for the directional Hubble factors is required.}
$\theta_i\, \theta_j$ to write 
\begin{equation}\label{BI:Hgop}
	\hat H_{\mathrm{g}} 
		= - \frac{1}{2 \kappa} \sum_{i \ne j} \abs{\hat v}^{\frac{1 + n}{4}}\, \hat\theta_i\, \abs{\hat v}^{\frac{1 + n}{2}}\, \hat\theta_j\, \abs{\hat v}^{\frac{1 + n}{4}}, 
\end{equation}
where $\bra{\vec\lambda} \hat v := v_0 \Lambda \bra{\vec\lambda}$ denotes the \emph{signed} volume operator and $v_0 := \abs{p_0}^{3/2}$ is the natural quantum of volume.  This expression uses a state-dependent lapse $N$ proportional to the $n^{\text{th}}$ power of the volume.  The harmonic time gauge used in \cite{aw2009} corresponds to $n = 1$.

\section{Reduction to the Isotropic Model}

\subsection{Reduction of the Symmetry Constraints}
\label{SR:Constraints}

The companion paper \cite{behm2017} selects the homogeneous and isotropic section of general relativity by setting 
\begin{equation}\label{FullSC}
	\mbb{S}[f, g] := \mbb{B}[f]\, \vol[g] - \vol[f]\, \mbb{B}[g] \approx 0, 
\end{equation}
for arbitrary smearing fields $f_{ij}$ and $g_{ij}$, where 
\begin{equation}\label{FullBV}
	\mbb{B}[f] := \sgn(\det e) \sum_{ij} \int_{\mathcal{V}} \mbb{F}^i\! \wedge e^j\, f_{ij} 
	\qquad\text{and}\qquad
	\vol[f] :=  \int_{\mathcal{V}} \tr f\, \abs{\det e}.
\end{equation}
The curvature appearing in the definition of $\mbb{B}[f]$ is that of the complexified connection 
\begin{equation}\label{FullA}
	\mbb{A}_a^i := A_a^i + \ii \alpha \, e_a^i,
\end{equation}
where $\alpha$ is an arbitrary, but fixed, real constant with units of inverse length.  The conditions (\ref{FullSC}) imposing homogeneity and isotropy in this approach are diffeomorphism covariant in the sense that replacing both the fundamental fields $(A, E)$ and the smearing fields $(f, g)$ with their images under a spatial diffeomorphism leaves $\mbb{S}[f, g]$ unchanged.  Requiring (\ref{FullSC}) for \emph{all} choices of the smearing fields therefore selects those points $(A, E)$ of the phase space that are invariant under \emph{some} action, as opposed to under a \emph{fixed} action, of one of the symmetry groups appropriate for isotropic and homogeneous cosmologies.

The symmetry conditions (\ref{FullSC}) simplify considerably when restricted to the phase space of Bianchi I cosmologies described in the previous section.  Specifically, (\ref{FullA}) becomes 
\begin{align*}
	\mbb{A}^i 
		= \mbb{c}^i\, \frac{\mathring{e}^i}{L_i} 
		:= \biggl( c^i + \ii\alpha\, \frac{\vol(\vec p)}{p_i} \biggr) \, \frac{\mathring{e}^i}{L_i}, 
\end{align*}
and the functionals from (\ref{FullBV}) become 
\begin{align*}
	\mbb{B}[f] = \sum_i \frac{p_x\, p_y\, p_z\, \mbb{c}^x\, \mbb{c}^y\, \mbb{c}^z}{p_i\, \mbb{c}^i \vol^2(\vec p)} \int_{\mathcal{V}} f_{ii}\, \abs{\det e} 
	\qquad\text{and}\qquad 
	\Vol[f] = \sum_i \int_{\mathcal{V}} f_{ii}\, \abs{\det e}, 
\end{align*}
respectively.  These both depend on the smearing field $f_{ij}$ only through the average values 
\begin{align*}
	f_i := \frac{1}{\vol(\vec p)} \int_{\mathcal{V}} f_{ii}\, \abs{\det e} 
\end{align*}
of its diagonal components over the fiducial cell.  Note that such an average is independent of the (homogeneous) triad field in a Bianchi I geometry.  Using these averages, together with the definition (\ref{BI:dHub}) of the directional Hubble rates as functions on phase space, then gives 
\begin{equation}\label{BI:Bdef}
	\mbb{B}[f] 
		= \sum_i f_i\, \mbb{B}^i 
		:= \vol(\vec p) \sum_i f_i \Biggl( \prod_{j \ne i} (\bip\theta_j + \ii\alpha) \Biggr)
	\qquad\text{and}\qquad 
	\vol[f] = \vol(\vec p) \sum_i f_i.
\end{equation}
Finally, substituting these expressions into (\ref{FullSC}) gives the symmetry conditions 
\begin{equation}\label{BI:redSC}
	\mbb{S}[f, g] = \vol(\vec p) \sum_{ij} f_i\, g_j\, (\mbb{B}^i - \mbb{B}^j) \approx 0
\end{equation}
that select the isotropic sector of the Bianchi I model.

The content of the reduced symmetry conditions becomes clearer if we rewrite $\mbb{B}^i - \mbb{B}^j = \mbb{B}^i\, \varsigma^j - \varsigma^i\, \mbb{B}^j$ in the sum, where $\varsigma^i$ denotes the vector with all components equal to one.  Doing so shows that the symmetry conditions hold for all smearing fields if and only if $\mbb{B} \wedge \varsigma = 0$, meaning that the two vectors are proportional, and thus that $\mbb{B}^x = \mbb{B}^y = \mbb{B}^z$.  Furthermore, we have that 
\begin{equation}\label{SSBPB}
	(\bip\theta_y + \ii\alpha) (\bip\theta_z + \ii\alpha) 
		= (\bip\theta_x + \ii\alpha) (\bip\theta_z + \ii\alpha) 
		= (\bip\theta_y + \ii\alpha) (\bip\theta_z + \ii\alpha) 
	\qquad\Leftrightarrow\qquad 
	\theta_x = \theta_y = \theta_z.
\end{equation}
Thus, the full content of the symmetry conditions (\ref{FullSC}) in the Bianchi I model is just that all three directional Hubble rates are the same.

In order to impose the symmetry conditions (\ref{BI:redSC}) simultaneously in the canonical formalism, one must check that their Poisson algebra closes.  To do so, first define 
\begin{equation}\label{SR:scred}
	\mbb{S}[f, g] = \sum_{ijk} \epsilon^{ijk}\, f_i\, g_j\, \mbb{S}_k 
	\qquad\text{with}\qquad 
	\mbb{S}_k := \vol(\vec p) \sum_{lm} \epsilon_{klm}\, \mbb{B}^l.
\end{equation}
The symmetry conditions, for all smearing fields $f_{ij}$ and $g_{ij}$, is equivalent to $\mbb{S}_x = \mbb{S}_y = \mbb{S}_z = 0$ due to the homogeneity of the Bianch I model.  Furthermore, 
\begin{align}
	\pbrack[\big]{\mbb{S}_x}{\mbb{S}_y} 
		&= \pbrack[\Big]
				{\bigl( c^x p_x + \ii\alpha \vol(\vec p) \bigr) \bigl( c^y p_y - c^z p_z \bigr)}
				{\bigl( c^y p_y + \ii\alpha \vol(\vec p) \bigr) \bigl( c^z p_z - c^x p_x \bigr)}
			\notag\\[1ex]
		&= \bigl( c^x p_x + \ii\alpha \vol(\vec p) \bigr) \bigl( c^z p_z - c^x p_x \bigr) \pbrack[\Big]
				{c^y p_y - c^z p_z}
				{\ii\alpha \vol(\vec p)}
			\notag\\&\hspace{2em}
			+ \bigl( c^y p_y + \ii\alpha \vol(\vec p) \bigr) \bigl( c^y p_y - c^z p_z \bigr) \pbrack[\Big]
				{\ii\alpha \vol(\vec p)}
				{c^z p_z - c^x p_x}
			\notag\\&\hspace{2em}
			+ \bigl( c^y p_y - c^z p_z \bigr) \bigl( c^z p_z - c^x p_x \bigr) \Bigl( 
				\pbrack[\big]{c^x p_x}{\ii\alpha \vol(\vec p)}
				+ \pbrack[\big]{\ii\alpha \vol(\vec p)}{c^y p_y} \Bigr)
			\notag\\[1ex]
		&= 0, 
\end{align}
and cyclic permutations.  We have used $\pbrack{c^i p_i}{c^j p_j} = 0$ in passing to the second line here, as well as $\pbrack{c^i p_i}{\vol(\vec p)} = \frac{1}{2} \kappa\bip \vol(\vec p)$ in the final step.  This is a stronger result than in the full theory \cite{behm2017}, where the Poisson algebra of the symmetry conditions is closed (\textit{i.e.}, the Poisson bracket of two $\mbb{S}'s$ is a sum of terms proportional to $\mbb{S}$'s) but not trivial.  A similar calculation shows that 
\begin{equation}
	\pbrack[\big]{\mbb{S}_x}{\bar{\mbb{S}}_y} 
		= - \ii\alpha\kappa\bip\, \bigl( c^y p_y - c^z p_z \bigr) \bigl( c^z p_z - c^x p_x \bigr) \vol(\vec p). 
\end{equation}
Although this Poisson bracket does not vanish everywhere in phase space, it does vanish when the symmetry conditions hold.  Again, this is a stronger result than in the full theory \cite{behm2017}, where the Poisson brackets of the symmetry conditions and their complex conjugates generally do not vanish even weakly, \textit{i.e.}, on the submanifold where the symmetry conditions hold.  
This result is attributable to the proportionality
between each symmetry condition and its complex conjugate in the homogeneous Bianchi I model with coefficient non-zero and smooth throughout $\Gamma$, 
\begin{equation}\label{eqn:SpropSbar}
	\bar{\mbb{S}}_y 
		= \bigl( c^y p_y - \ii\alpha\vol(\vec p) \bigr) \bigl( c^z p_z - c^x p_x \bigr) 
		= \frac{c^y p_y - \ii \alpha\vol(\vec p)}{c^y p_y + \ii\alpha\vol(\vec p)}\, \mbb{S}_y
		=: \eta_y \mathbb{S}_y,
\end{equation}
as it implies immediately
\begin{align}
\label{eqn:SSbarfirst}
    \{\mathbb{S}_x, \bar{\mathbb{S}}_y\} = \{\mathbb{S}_x, \eta_y\} \mathbb{S}_y + \{\mathbb{S}_x, \mathbb{S}_y\} \eta_y \approx 0 .
\end{align}

\subsection{Reduction to the Classical Isotropic Sector}

Recall $\Gamma \cong \Re^3 \times \Re_{\!\mathord\times}^3$ denotes the classical phase space of the Bianchi I model constructed in the previous section.  Let $\bar\Gamma \subset \Gamma$ denote the \defn{classical isotropic sector} on which the symmetry conditions (\ref{FullSC}) hold, or equivalently, on which $\theta_x = \theta_y = \theta_z$.  There are only two independent conditions here, so $\bar\Gamma$ is (locally) a 4-dimensional submanifold of the 6-dimensional phase space $\Gamma$.  We can pull the symplectic structure (\ref{BI:redss}) back to $\bar\Gamma$ by first writing 
\begin{align*}
	\Omega 
		= \frac{1}{\kappa\bip} \sum_i \ed c^i \wedge \ed p_i 
		= \frac{1}{\kappa} \sum_i \ed \biggl( \frac{\theta_i \vol(\vec p)}{p_i} \biggr) \wedge \ed p_i
		= \frac{1}{\kappa}\, \ed \Biggl( \vol(\vec p) \sum_i \theta_i\, \ed\ln |p_i| \Biggr)
\end{align*}
in complete generality.  If we now set $\theta_x = \theta_y = \theta_z =: \theta$, then 
\begin{equation}\label{SR:ssred}
	\ed\ln\vol(\vec p) = \frac{1}{2} \sum_i \ed \ln |p_i| 
	\qquad\leadsto\qquad
	\underline{\Omega} = \frac{2}{\kappa}\, \ed\theta \wedge \underline{\ed\vol(\vec p)}.
\end{equation}
This is clearly degenerate, with a kernel consisting of vectors tangent to $\bar\Gamma \subset \Gamma$ that change neither the common value $\theta$ of the directional Hubble rates, nor the proper volume $\vol(\vec p)$ of the fiducial cell.

The appropriate, non-degenerate \defn{isotropic phase space} is the quotient manifold $\Gamma\!_S \cong \Re \times \Re_{\!\mathord\times}$, consisting of equivalence classes of points $(c^i, p_i) \in \bar\Gamma$ on which the geometric means 
\begin{align*}
	c := \sqrt[3]{c^x c^y c^z} 
	\qquad\text{and}\qquad
	p := \sqrt[3]{p_x p_y p_z} 
\end{align*}
both take constant values.  It is sometimes convenient to use the \emph{signed} volume 
\begin{align*}
	v := \sgn p\, |p|^{3/2}
\end{align*}
instead of $p$ itself as a phase-space coordinate for the isotropic model.  The volume and Hubble rate(s) are 
\begin{equation}\label{SR:VHdef}
	\underline{\vol(\vec p)} = |p|^{3/2} = \abs{v}
	\qquad\text{and}\qquad 
	\theta 
		= \sqrt[3]{\smash{\underline\theta_x \underline\theta_y \underline\theta_z}\vphantom{\theta_x}}
		= \frac{c \sgn p}{\bip\, |p|^{1/2}}
		= \frac{c}{\bip v^{1/3}}
\end{equation}
respectively, on each equivalence class in $\Gamma\!_S$.  They therefore descend to well-defined functions on the reduced phase space.  Any function of these quantities likewise descends to $\Gamma\!_S$, including in particular the gravitational part (\ref{BI:Hred}) of the Hamiltonian constraint 
\begin{equation}\label{SR:iHgop}
	\underline{\Hg} 
		= - \frac{3}{\kappa\bip^2}\, |p|^{(3 n + 1)/2}\, c^2
		= - \frac{3}{\kappa\bip^2}\, |v|^{n + 1/3}\, c^2, 
\end{equation}
where we have fixed the lapse $N = \vol(\vec{p})^n$, as well as the regularized Hubble rates (\ref{BI:rhub}) 
\begin{align*}
	\underline{\bar\theta_i(s)}
		= \frac{\sgn p}{\bip s} \sin \frac{c s}{|p|^{1/2}} 
		= \frac{1}{\bip s} \sin \frac{c s}{v^{1/3}},
\end{align*}
and the regularized Hamiltonian constraint (\ref{BI:Hreg}) derived from them.  The symplectic structure (\ref{SR:ssred}) becomes 
\begin{align*}
	\underline{\Omega} 
		= \frac{3}{\kappa\bip}\, \ed c \wedge \ed p 
		= \frac{2}{\kappa\bip\, |v|^{1/3}}\, \ed c \wedge \ed v
		= \frac{2}{\kappa} \ed(\theta \sgn p) \wedge \ed v .
\end{align*}
This clearly descends to $\Gamma\!_S$ as well, where it is equivalent to the standard Poisson bracket $\pbrack{c}{p} = \frac{1}{3} \kappa\bip$ of isotropic loop quantum cosmology.

As mentioned in the previous section, the residual automorphism group for the isotropic model is naturally isomorphic to the center $Z(\operatorname{Aut}_R)$ of the residual automorphism group for the Bianchi I model.  In detail, the isotropic dilatations (\ref{BI:idil}) and isotropic reflections (\ref{BI:ipar}) act on $\Gamma\!_S$ via 
\begin{equation} \label{SR:isoaut}
	\varphi_T : (c, p) \mapsto \bigl( \ee^{-T/3}\, c, \ee^{-2T/3}\, p \bigr) 
	\qquad\text{and}\qquad 
	\varphi_Z : (c, p) \mapsto \bigl( Z c, Z p \bigr), 
\end{equation}
respectively.  More importantly, however, the complementary subgroup $\operatorname{Aut}_R^0 \cong \operatorname{Aut}_R / Z(\operatorname{Aut}_R)$ of proper residual automorphisms  acts \emph{transitively} on the equivalence class of points $(c^i, p_i) \in \bar\Gamma$ corresponding to any given point $(c, p) \in \Gamma\!_S$ of the isotropic phase space.  To see this, first observe that 
\begin{align*}
	(c^i, p_i) 
		= \biggl( \frac{c p}{\theta}\, \frac{\theta_i}{p_i}, \frac{p_i}{p}\, p \biggr) 
		= \varphi_{\vec t} \circ \varphi_{\vec\zeta} \biggl( \frac{c}{\theta}\, \theta_i, (p, p, p) \biggr)
	\qquad\text{with}\qquad 
	\left\{ \begin{aligned}
		t_i &:= \ln |p_i| - \ln |p| \\
		\zeta_i &:= \sgn(pp_i)
	\end{aligned} \right.
\end{align*}
for any point $(c^i, p_i) \in \Gamma$, where $\theta := \sqrt[3]{\theta_x \theta_y \theta_z}$ denotes the geometric mean of the directional Hubble rates.  The residual automorphism on the right here is proper because $\sum_i t_i = 0$ and $\zeta_x \zeta_y \zeta_z = 1$ by construction.  Inverting it shows that \emph{every} $(c^i, p_i) \in \Gamma$ can be put in a ``partly diagonal'' form with $p_x = p_y = p_z = p$ by an appropriate proper residual automorphism.  Furthermore, the resulting phase-space point is ``fully diagonal'' in the sense that $c^x = c^y = c^z = c$ as well if and only if $(c^i, p_i) \in \Gamma$ lies in the classical isotropic sector $\bar\Gamma \subset \Gamma$ where $\theta_x = \theta_y = \theta_z = \theta$.  This fact characterizes the classical isotropic sector $\bar\Gamma \subset \Gamma$ purely in terms of the action of the residual automorphism group: 
\begin{theorem}\label{thm:sym}
A point $(c^i, p_i) \in \Gamma$ of the Bianchi I phase space lies in the classical isotropic sector $\bar\Gamma \subset \Gamma$ if and only if there exists a residual automorphism $\varphi \in \operatorname{Aut}_R$ such that 
\begin{equation}\label{eq:rho}
	\varphi \circ \varphi_\pi \circ \varphi^{-1} (c^i, p_i) = (c^i, p_i)
\end{equation}
for all residual rotations $\varphi_\pi \in \operatorname{Rot}_R$.  One may choose $\varphi \in \operatorname{Aut}_R^0$ to be proper without loss of generality.
\end{theorem}

\subsection{Quantum Isotropy and the Isotropic Model}
\label{Iso:Quantum}

Working in the Hilbert space $\mfs{H}$ of the Bianchi I model, we define the (regularized) operator analogues of the functions $\mbb{S}_i$ from (\ref{SR:scred}) that define the classical isotropic sector as follows: 
\begin{align}\label{SCReQ}
    \hat{\mbb{S}}_x(s) = |\hat v|^{\frac{1}{2}}\, \frac{
        \bigl( \bip\hat\theta_x(s) + \ii\alpha \bigr)\, |\hat v|\, \bigl( \bip\hat\theta_z(s) - \bip \hat\theta_y(s) \bigr) 
        + \bigl( \bip\hat\theta_z(s) - \bip \hat\theta_y(s) \bigr)\, |\hat v|\, \bigl( \bip\hat\theta_x(s) + \ii\alpha \bigr)}{2}\, |\hat v|^{\frac{1}{2}}, 
\end{align}
and cyclic permutations.  The regularized Hubble rate operators $\hat\theta_i(s)$ are defined in  (\ref{BI:dHops}), and the ordering prescription adopted here at the quantum level mimics that of the Hamiltonian constraint from (\ref{BI:Hgop}).  The \defn{quantum isotropic sector} is the subspace  $\mathcal{V}_{\text{symm}} \subset \mfs{H}$ of Bianchi I states that are annihilated by all three operators $\hat{\mbb{S}}_i$.  It is not obvious at the moment that any such states exist.  But we will see in the next section that indeed they do by showing that all three operators annihilate every state in a particular embedding of the Hilbert space of the fully isotropic theory into $\mfs{H}$.

To compare the isotropic sector of the quantum Bianchi I model to the quantum isotropic model --- wherein isotropy is imposed at the classical level, prior to quantization --- we must of course review the quantization of the fully reduced model itself.  It proceeds \cite{aps2006} similarly to that of the Bianchi I model presented in detail above.  We introduce the the exponentials 
\begin{align*}
\Delta(s) := \exp -\frac{ic\, s}{2|p|^{1/2}}, 
\end{align*}
and motivate their quantization by recalling that $c$ becomes a differential operator in a Schr\"odinger quantization based on $p$, and 
\begin{align*}
    \hat{\Delta}(s) := \exp \frac{s\hbar \kappa \gamma}{6|p|^{1/2}}\frac{d}{dp}
    = \exp \frac{s\hbar\kappa \gamma}{4}\frac{d}{dv}.
\end{align*}
a shift operator acting on $\hat{v}$ eigenstates as
\begin{align*}
\langle v | \hat{\Delta}(s) = \bra[\Big]{v + \frac{s\hbar\kappa \gamma}{4}} = \bra[\Big]{v + \frac{1}{2} \frac{s}{s_0} v_0}.
\end{align*}
The Hubble rate (\ref{SR:VHdef}) can again be expressed
as limits of combinations of $\Delta(s)$ and $v$:
\begin{align*}
\theta = \lim_{s \rightarrow 0} \theta(s) := \lim_{s \rightarrow 0}
\frac{\sgn v (\Delta(-2s) - \Delta(2s))}{2i\gamma s} .
\end{align*}
Weyl ordering yields the regulated operator
\begin{align}
\nonumber
\hat{\theta}(s)
=& \frac{\sgn \hat{v} (\hat{\Delta}(-2s) - \hat{\Delta}(2s)) 
+ (\hat{\Delta}(-2s) - \hat{\Delta}(2s))\sgn \hat{v} }{4i\gamma s} \\
=&
\Bigl(\hat\Delta(-2s) \,
			\Theta \big( \abs[\nml]{\hat{v}} + (v_0 s/s_0)\sgn\hat v\bigr)
			- \hat\Delta(2s)\,
			\Theta \bigl( \abs[\nml]{\hat{v}} - (v_0 s/s_0)\sgn\hat v \bigr) \Bigr) \frac{\sgn\hat v}{2 \ii \bip s}.
\label{Iso:dHops}
\end{align}
Following \cite{aps2006}, we again take the limits to $s=s_0$,
so that $\hat{\theta} = \hat{\theta}(s_0)$. 
As in the Bianchi I case, the isotropic Hamiltonian constraint (\ref{SR:iHgop}) can be expressed in terms of $v$ and $\theta$ 
and quantized using a symmetric ordering, yielding the constraint operator of \cite{aps2006},
\begin{align}\label{SR:Hgop}
\hat{H}_g = - \frac{3}{\kappa} |\hat{v}|^{\frac{1+n}{4}}
\hat{\theta} |\hat{v}|^{\frac{1+n}{2}} \hat{\theta}
|\hat{v}|^{\frac{1+n}{4}} .
\end{align}

\section{Embedding}

\subsection{Conditions on, and Desired Properties of, the Embedding}

\def\Full{{\rm Full}}

The definition of a gauge- and diffeomorphism-invariant homogeneous isotropic sector in full loop quantum gravity is only the first part of the strategy outlined in \cite{behm2017}. The second part is to define an embedding $\iota_\Full$ of the isotropic model into this sector, and use this embedding to compare operators and dynamics in the two models. The strategy presented there is to define $\iota_\Full$ by stipulating the following conditions:
\begin{enumerate}
\item[(i)] $\iota_\Full$ should map states into the quantum homogeneous isotropic sector.
That is, it's image should be annihilated by the symmetry constraint operators $\hat{\mbb{S}}[f,g]$ for all $f$ and $g$.
\item[(ii)] $\iota_\Full$ should \textit{intertwine} two pairs of operators
$(\hat{O}^i_\Full, \hat{O}^i_S)$, $i=1,2$ in the full and homogeneous isotropic theories,
\begin{align}
\label{full:intertwine}
    \hat{O}^i_\Full \circ \iota_\Full = \iota_\Full \circ \hat{O}^i_S ,
\end{align}
corresponding to the two dimensions of the homogeneous isotropic phase space.
\end{enumerate}
The first condition fixes the image of $\iota_\Full$,
while the second condition fixes how states in this image are identified with states in the homogeneous isotropic model.
If we use $\iota_\Full$ to identify homogeneous isotropic states with full theory states, the second condition (\ref{full:intertwine}) simply states that
\textit{$\hat{O}^i_\Full$ should have the same action on homogeneous isotropic states as $\hat{O}^i_S$}.

For the present paper, the task is to find an embedding $\iota$ of the isotropic quantum model into the \mbox{Bianchi I} quantum model.  The analogue of the above conditions is then
\begin{enumerate}
\item[(i)] $\hat{\mathbb{S}}_i \circ \iota = 0$ for all $i$.
\item[(ii)] $\hat{O}^i \circ \iota = \iota \circ \hat{O}^i_S$
for two pairs of operators $(\hat{O}^i, \hat{O}^i_S)$
in the \mbox{Bianchi I} and isotropic models, $i=1,2$.
\end{enumerate}
In contrast to full loop quantum gravity, in the \mbox{Bianchi I} model there are automorphisms with well-defined action in the quantum theory that are non-trivial \textit{even once the Gauss and diffeomorphism constraints are imposed}. 
%
%Classically, in full canonical gravity, one can divide diffeomorphisms into two types: Those that approach the identity at infinity, and those that do not.
%The former are generated by the diffeomorphism constraint, so that imposition of the quantum diffeomorphism constraint enforces invariance of all LQG states under such diffeomorphisms. But the spin-networks which span the LQG state space all represent geometries of \textit{finite extent}, and hence none of them can distinguish between diffeomorphisms that do and do not approach the identity at infinity. Hence they are also invariant under diffeomorphisms that do not approach identity at infinity. In the \mbox{Bianchi I} case, by contrast, due to homogeneity, all states necessarily represent geometries with infinite extent, and hence \textit{can} distinguish between these two types of diffeomorphisms. As a consequence, even though the diffeomorphism constraint is identically satisfied in the \mbox{Bianchi I} model, there are still diffeomorphisms with a non-trivial action in model, what we have called above the `residual diffeomorphisms' in section \ref{}. Among these, only those that preserve the symplectic structure --- namely the volume preserving residual diffeomorphisms --- have a corresponding action on quantum states. 
%
%
As a consequence, in the \mbox{Bianchi I} case, there is an additional covariance condition which can and must be stipulated:
\begin{enumerate}
\item[(iii)] $\iota$ should be covariant under all residual automorphisms well-defined in the quantum theory.
\end{enumerate}
%Keep as comment-note only: 
%While \textit{covariance} of $\iota$ under residual diffeos/automorphisms is 
%is physically justified, imposing \textit{invariance}under residual diffeos/automorphisms on the isotropic sector is not: The generators of the canonical residual diffeos are not constraints, but rather, their vanishing is equivalent to our isotropy condition.
As we shall argue below,
conditions (i) and (iii) are expected to have the same content from classical analysis, and, in the quantum theory, we will see that (iii) implies (i). 
For this reason, we impose (iii), and let (i) follow as a 
consequence.\footnote{Once a single superselection sector is picked in the \mbox{Bianchi I} model \cite{aw2009}, the implication also goes in the opposite direction.
However, the argument for superselection comes from a specific dynamics. 
Part of the purpose of this work is to test compatibility of the dynamics in the isotropic and \mbox{Bianchi I} models, so that we preferred our presentation to be independent of any one choice of dynamics, and hence independent of any superselection.}
In fact, the classical analysis will lead us to expect not only the equivalence of (i) and (iii), but also the equivalence of
\begin{enumerate}
    \item[$\overline{\text{(i})}$]
    $\hat{\mathbb{S}}_i^\dagger \circ \iota = 0$
\end{enumerate}
with both of these, and we will see explicitly in the quantum theory that (iii) implies not only (i), but $\overline{\text{(i)}}$ as well. 

In this section, our imposition of (iii) ---  basically equivalent to (i) --- and (ii) will uniquely determine $\iota$. This is consistent with the results found for 
the toy model in appendix B of \cite{behm2017}. 
Once uniquely determined, $\iota$ can be used to compare other operators $(\hat{O}, \hat{O}_s)$ in the two models, again via the intertwining condition
\begin{align}
    \hat{O} \circ \iota_\Full = \iota_\Full \circ \hat{O}_S .
\end{align}
Note that if $\hat{O}_S$ is not known, the above equation will also uniquely determine it. Hence, the above equation can also be thought of as defining a \textit{map} from \mbox{Bianchi I} operators preserving the isotropic sector, to operators in LQC. 
Remarkably, in the end, we will find that $\iota$ maps \textit{all} of the physically relevant operators in \mbox{Bianchi I} introduced in section \ref{BI:Quantum} exactly to the corresponding operators in the isotropic theory introduced in section \ref{Iso:Quantum}. 
This includes the Hamiltonian constraint operators in the two models, so that the embedding $\iota$ will establish that the isotropic model captures both the kinematics and dynamics of the isotropic sector of the quantum \mbox{Bianchi I} \textit{exactly}.

\subsection{Unitary Action of Canonical Residual Automorphisms}
\label{ss:unitaryaut}

Let ${\rm Dil}_R^o$ denote the \textit{\textbf{proper anisotropic dilatations}}, that is, the dilatations preserving the volume of the fiducial cell.
The subgroup of the residual automorphisms introduced in subsection \ref{BI:Classical} that are canonical transformations, hence with unitary action on quantum states, we call the \textit{\textbf{canonical residual automorphisms}} ${\rm Aut}^C_R$. Explicitly, it is generated by 
the proper anisotropic dilatations, the partial reflections, and the residual rotations, ${\rm Aut}^C_R = ({\rm Dil}^o_R \times {\rm Par}_R) \rtimes {\rm Rot}_R$.
From equations (\ref{BI:anis}), (\ref{BI:pref}), 
and (\ref{BI:rrot}), for each $\dilparam_x, \dilparam_y \in \mathbb{R}$, $\vec{\zeta} \in \{\pm 1\}^3$, and $\pi \in S_3$,
the actions of these three types of transformations in the quantum theory is given by 
\begin{align}
    \hat{\varphi}_{(\dilparam_x, \dilparam_y, -\dilparam_x-\dilparam_y)}& |p_x, p_y, p_z\rangle = |e^{\dilparam_x} p_x, e^{\dilparam_y} p_y, e^{-\dilparam_x-\dilparam_y} p_z\rangle, \notag \\
    \hat{\varphi}_{\vec{\zeta}} & |(p_i)\rangle = |(\zeta_i p_i)\rangle, \\
    \hat{\varphi}_\pi & |(p_i)\rangle = |(p_{\pi(i)})\rangle . \notag
\end{align}
As discussed in section \ref{s:BIrev}, the residual automorphisms, when acting on the isotropic phase space, reduce to the group of isotropic automorphisms ${\rm Aut}_S$. For canonical residual automorphisms $T:= \sum_{i=1}^3 t_i = 0$, so that 
${\rm Aut}^C_R$ reduces to the even smaller group of isotropic reflections ${\rm Par}_S$.
That is, the actions of the proper anisotropic dilatations and residual rotations on the isotropic phase space are trivial,
while the action of the partial reflections is given by (\ref{SR:isoaut}), so that the quantum action is given by
\begin{align}
    \hat{\varphi}_{(\dilparam_x, \dilparam_y, -\dilparam_x-\dilparam_y)} |p\rangle &= |p\rangle, \notag \\
    \hat{\varphi}_{\vec{\zeta}} |p\rangle &= |\zeta_1 \zeta_2 \zeta_3 p \rangle, \\
    \hat{\varphi}_\pi |p\rangle &= |p\rangle . \notag
\end{align}

\subsection{Derivation}

\paragraph{Imposition of Covariance under Residual Canonical Automorphisms}
% \subsubsection{Imposition of Covariance under Residual Canonical Automorphisms}
% \label{covsec}

For the purpose of deriving the embedding, it is convenient to label the momentum basis in \mbox{Bianchi I} using $\lambda_x$, $\lambda_y$, and $\Lambda$ (\ref{BI:affp}, \ref{BI:braD}), and to label the momentum basis in the isotropic theory also by $\Lambda$. In terms of these labels, the action of the most general canonical residual automorphism 
$\varphi := \varphi_{(t_x,t_y,-t_x-t_y)} \circ
\varphi_{\vec{\zeta}} \circ \varphi_\pi$ is given by
\begin{align*}
\hat{\varphi} |\lambda_x, \lambda_y, \Lambda \rangle &= \ket[\Big]{\zeta_x e^{\dilparam_x/2} \lambda_{\pi(x)},\, \zeta_y e^{\dilparam_y/2} \lambda_{\pi(y)}, \, \zeta_x \zeta_y \zeta_z \Lambda} \\
\hat{\varphi} |\Lambda \rangle 
&= |\zeta_x \zeta_y \zeta_z \Lambda \rangle .
\end{align*}
Imposing covariance of $\iota$ under all such transformations,
$\hat{\varphi} \circ \iota = \iota \circ \hat{\varphi}$,
leads to the following condition on the matrix elements of $\iota$:
\begin{align}
\label{eq:fullcov}
\langle \lambda_x, \lambda_y, \Lambda | \iota | \Lambda' \rangle
= \langle \zeta_x e^{-\dilparam_x/2} \lambda_{\pi(x)}, \zeta_y e^{-\dilparam_y/2}\lambda_{\pi(y)}, \zeta_x \zeta_y \zeta_z \Lambda | \iota | \zeta_x \zeta_y \zeta_z \Lambda' \rangle
\end{align}
for all $t_x, t_y, \vec{\zeta}, \pi$. First setting $\zeta_z = \zeta_x \zeta_y$ and $\pi = {\rm id}$, imposing this condition for all $t_x, t_y, \zeta_x, \zeta_y$ leads to
\begin{align*}\langle \lambda_x, \lambda_y, \Lambda | \iota | \Lambda' \rangle
= \langle \beta_x \lambda_{x}, \beta_y \lambda_{y}, \Lambda | \iota | \Lambda' \rangle 
\end{align*}
for all $\beta_x, \beta_y \in \mathbb{R}_{\!\times}$. 
Since $\lambda_x \neq 0$ and $\lambda_y \neq 0$, 
setting $\beta_x = \lambda_x^{-1}$ and $\beta_y = \lambda_y^{-1}$, we have
\begin{align*}\langle \lambda_x, \lambda_y, \Lambda | \iota | \Lambda' \rangle
= \langle 1, 1, \Lambda | \iota | \Lambda' \rangle 
=: C(\Lambda; \Lambda') 
\end{align*}
for all $(\lambda_x, \lambda_y, \Lambda)\in \mathbb{R}_{\!\times}^3$.
These matrix elements then furthermore satisfy (\ref{eq:fullcov}) for \textit{all} canonical residual automorphisms if and only $C(\Lambda; \Lambda')$ additionally satisfies $C(-\Lambda; -\Lambda') = C(\Lambda; \Lambda')$. Explicitly, the resulting embedding then takes the form
\begin{align}
\label{eq:coviota}
%\begin{split}
\iota |\Lambda' \rangle = \sum_{\lambda_x, \lambda_y, \Lambda \neq 0} C(\Lambda; \Lambda') 
|\lambda_x, \lambda_y, \Lambda \rangle 
&= \sum_{\Lambda \neq 0} C(\Lambda; \Lambda') \sum_{\lambda_x, \lambda_y \neq 0}
|\lambda_x, \lambda_y, \Lambda \rangle 
=:
 \sum_{\Lambda \neq 0} C(\Lambda; \Lambda') \iota_0
|\Lambda \rangle .
%\end{split}
\end{align}
Note that, from this condition alone, 
every non-zero element in the range of $\iota$ is \textit{necessarily non-normalizable} in the polymer inner product (\ref{BI:polyip}). Define $\Cyl_{BI}$ to be the space of finite linear combinations of momentum eigenstates in the \mbox{Bianchi I} Hilbert space. Then, 
what we are saying is that covariance under canonical residual automorphisms forces 
the image of $\iota$ 
to be represented in the algebraic dual $\Cyl_{BI}^*$, which includes possibly non-normalizable linear combinations of momentum eigenstates.
This is similar to the kinematical non-normalizability of
diffeomorphism-invariant states in the full theory \cite{almmt1995}.

\paragraph{The Image of $\iota$ then Automatically Satisfies Quantum Isotropy Independent of Ordering Ambiguity.}
% \subsubsection{The Image of $\iota$ then Automatically Satisfies Quantum Isotropy Independent of Ordering Ambiguity.}
% \label{sect:qisosat}

\begin{lemma}
\label{lem:iota0int}
$\iota_0$ intertwines both $\hat{v}$ and $\hat{\theta}_i(s)$ for all $s$.
\end{lemma}
{\startproof

That $\iota_0$ intertwines $\hat{v}$ is immediate:
\begin{align*}
\iota_0 \circ \hat{v} |\Lambda \rangle = v_0 \Lambda \iota_0 |\Lambda \rangle
= v_0 \Lambda \sum_{\lambda_x, \lambda_y \neq 0}
|\lambda_x, \lambda_y, \Lambda \rangle
= \hat{v} \sum_{\lambda_x, \lambda_y \neq 0}
|\lambda_x, \lambda_y, \Lambda \rangle
= \hat{v} \circ \iota |\Lambda \rangle .
\end{align*}
For the $\hat{\theta}_i(s)$, it is sufficient to consider $\hat{\theta}_z(s)$.
Starting from equations (\ref{BI:dHops}) and (\ref{BI:braDs}),
we have for all $|\Lambda\rangle$,
\newcommand{\thisindent}{\dummy\hspace{0.5em}}
\begin{align*}
&\hat{\theta}_z(s) \circ \iota_0 |\Lambda \rangle
= 
\bigg[ \Delta_i(-2s) \,
\Theta\left(\left|\hat{\Lambda}\right|+(s/s_0)\sgn \hat{\lambda}_z\right)
- \Delta_i(2s)\,\Theta\left(\left|\hat{\Lambda}\right|-(s/s_0)\sgn \hat{\lambda}_z\right)
\bigg] \frac{\sgn \hat{\lambda}_z}{2i\gamma s}
\sum_{\lambda_x, \lambda_y \neq 0} |\lambda_x, \lambda_y, \Lambda \rangle \\
&\thisindent = \frac{1}{2i\gamma s} 
\sum_{\lambda_x, \lambda_y \neq 0}
\bigg[ \sgn \left(\frac{\Lambda}{\lambda_x \lambda_y}\right) \, \Theta\left(\left|\Lambda\right|+(s/s_0)\sgn\left(\frac{\Lambda}{\lambda_x\lambda_y}\right)\right)
\ket[\Big]{\lambda_x, \lambda_y, \Lambda + 
(s/s_0) \sgn(\lambda_x\lambda_y)}\\
&\thisindent \hspace{6.6em} - \sgn \left(\frac{\Lambda}{\lambda_x\lambda_y}\right) \, \Theta\left(\left|\Lambda\right|-(s/s_0) 
\sgn\left(\frac{\Lambda}{\lambda_x\lambda_y}\right)\right)
\ket[\Big]{\lambda_x, \lambda_y, \Lambda 
- (s/s_0) \sgn(\lambda_x\lambda_y)}
\bigg]
\\
&\thisindent = \frac{\sgn \Lambda}{2i\gamma s} 
\sum_{\lambda_x, \lambda_y \neq 0}
\bigg[ \{ \sgn(\lambda_x \lambda_y\} \, 
\Theta\left(\left|\Lambda\right|+(s/s_0)
\{\sgn (\lambda_x\lambda_y)\}\sgn\Lambda\right)
\ket[\Big]{\lambda_x, \lambda_y, \Lambda + 
(s/s_0) \{\sgn(\lambda_x\lambda_y)\} }\\
&\thisindent \hspace{6.6em} +\{- \sgn (\lambda_x\lambda_y)\} \, \Theta\left(\left|\Lambda\right|+(s/s_0)\{- \sgn (\lambda_x\lambda_y)\}
\sgn \Lambda\right)
\ket[\Big]{\lambda_x \lambda_y, \Lambda 
+ (s/s_0)\{-\sgn(\lambda_x\lambda_y)\}}
\bigg]
\\
&\thisindent = \frac{\sgn \Lambda}{2i\gamma s} 
\sum_{\lambda_x, \lambda_y \neq 0}
\bigg[ \Theta\left(\left|\Lambda\right|+(s/s_0)\sgn \Lambda\right)
\ket[\Big]{\lambda_x, \lambda_y, \Lambda + 
s/s_0} 
-\Theta\left(\left|\Lambda\right|-(s/s_0)\sgn \Lambda \right)
\ket[\Big]{\lambda_x, \lambda_y, \Lambda 
-s/s_0}
\bigg]
\\
&\thisindent = \iota_0 \, \frac{\sgn \Lambda}{2i\gamma s} 
\bigg[ \Theta\left(\left|\Lambda\right|+(s/s_0)\sgn \Lambda\right)
\ket[\Big]{\Lambda + s/s_0} 
-\Theta\left(\left|\Lambda\right|-(s/s_0)\sgn \Lambda \right)
\ket[\Big]{\Lambda - s/s_0}
\bigg]
\\
&\thisindent = \iota_0 \, \theta(s)
\ket[\Big]{\Lambda} .
\end{align*}
In going from line 3 to line 4, we have used the fact that the first and second terms
are identical except that the signs in braces in the first term are all $\sgn(\lambda_x\lambda_y)$, whereas all those in the second term are 
$-\sgn(\lambda_x\lambda_y)$, so that in exactly one of the two terms these signs are all $+1$ and in the other they are $-1$.
\finishproof}

\begin{theorem}
\label{thm:iotasymm}
$\iota$ as given in (\ref{eq:coviota}) satisfies 
\begin{align*}
{}^{\alpha'} \hat{\mathbb{S}}_i(s) \circ \iota = 0
\end{align*}
for all choices of regularization parameter $s$ and all choices of complexification parameter $\alpha'$, and 
independent of the choice of coefficients $C(\Lambda;\Lambda')$.
\end{theorem}
{\startproof

From equation (\ref{SCReQ}), 
for all $|\Lambda\rangle$, we have
\begin{align*}
&{}^{\alpha'}\hat{\mathbb{S}}_x(s) \circ \iota |\Lambda \rangle
= \frac{\gamma}{2} |\hat{v}|^{1/2}
\left( (\gamma \hat{\hubb}_x(s) + i\alpha') |\hat{v}| 
(\hat{\hubb}_y(s) - \hat{\hubb}_z(s))
+ (\hat{\hubb}_y(s) - \hat{\hubb}_z(s)) |\hat{v}| (\gamma \hat{\hubb}_x(s) + i\alpha')\right) |\hat{v}|^{1/2} \cdot \\
&\dummy \hspace{4.8in}
 \cdot \left(\sum_{\Lambda' \in \mathbb{R}_{\!\times}} C(\Lambda'; \Lambda)  \iota_0 |\Lambda' \rangle \right) \\
&= \frac{\gamma}{2} \sum_{\Lambda' \in \mathbb{R}_{\!\times}} C(\Lambda'; \Lambda) 
|\hat{v}|^{1/2}
\left( (\gamma \hat{\hubb}_x(s) + i\alpha') |\hat{v}| 
(\hat{\hubb}_y(s) - \hat{\hubb}_z(s))
+ (\hat{\hubb}_y(s) - \hat{\hubb}_z(s)) |\hat{v}| (\gamma \hat{\hubb}_x(s) + i\alpha')\right) |\hat{v}|^{1/2} \circ \iota_0 |\Lambda' \rangle \\
&= \frac{\gamma}{2} \sum_{\Lambda' \in \mathbb{R}_{\!\times}} C(\Lambda'; \Lambda) \iota_0 \circ
|\hat{v}|^{1/2}
\left( (\gamma \hat{\hubb}(s) + i\alpha') |\hat{v}| (\hat{\hubb}(s) - \hat{\hubb}(s))
+ (\hat{\hubb}(s) - \hat{\hubb}(s)) |\hat{v}| 
(\gamma \hat{\hubb}(s) + i\alpha')\right) |\hat{v}|^{1/2} 
|\Lambda' \rangle \\
&= 0 
\end{align*}
whence ${}^{\alpha'} \hat{\mathbb{S}}_z(s) \circ \iota= 0$
for all $s$ and $\alpha'$.
Similarly, 
${}^{\alpha'} \hat{\mathbb{S}}_y(s) \circ \iota= 
{}^{\alpha'} \hat{\mathbb{S}}_z(s) \circ \iota=0$
for all $s$ and $\alpha'$.
\finishproof}

Note that if \textit{any other ordering} of $\hat{\mathbb{S}}^i$ had been chosen in (\ref{SCReQ}),
this theorem would still hold.
Furthermore, for the case $\alpha' = -\alpha$,
this theorem implies that not only $\hat{\mathbb{S}}_i \equiv {}^\alpha \hat{\mathbb{S}}_i(s_0)$ annihilates $\iota$,
\textit{but also its adjoint} $\hat{\mathbb{S}}_i^\dagger \equiv {}^{-\alpha}\hat{\mathbb{S}}^i(s_0)$.
This contrasts with the full theory analysis in \cite{behm2017},
where one only expects it to be possible for an embedding to be annihilated by one of
$\widehat{\mathbb{S}[f,g]}$, $\widehat{\overline{\mathbb{S}[f,g]}}$, not both. Thus, the condition satisfied by $\iota$ in the \mbox{Bianchi I} case is much stronger. The possibility of this was expected due to equation (\ref{SSBPB}) in the classical theory, 
and this will be discussed in section \ref{sec:origins}.

\paragraph{Consistency with Classical Theory}

It may seem puzzling that canonical residual automorphism covariance of $\iota$ implies that its image satisfies our quantum isotropy condition: Is not the former simply a condition of consistency with gauge symmetry, whereas the latter is an actual physical restriction on states? It may seem equally puzzling that it simultaneously implies that the adjoint of our isotropy condition is satisfied on the image of $\iota$.

These puzzles are resolved if one carefully translates these logical relations to the classical theory, where we will see that it holds as well. The canonical residual automorphism covariance of $\iota$ implies that the image of $\iota$ is invariant under 
the identity component of this group, the proper anisotropic dilatations. 
The classical analogue of imposing invariance under a unitary flow in quantum theory, $e^{t\hat{X}} |\Psi\rangle = |\Psi\rangle$,  is to impose that the corresponding \textit{generators} be zero: $\hat{X}|\Psi\rangle = 0 \leadsto X \approx 0$.
The proper anisotropic dilatations are the flows on space generated by vector fields of the form $\vec{X}_{\dilparam_x, \dilparam_y} := \dilparam_x x \frac{\partial}{\partial x} + \dilparam_y y \frac{\partial}{\partial y} - (\dilparam_x+\dilparam_y)z \frac{\partial}{\partial z}$. The corresponding 
canonical generators on the phase space are thus
\begin{align}
\begin{split}
X_{\dilparam_x, \dilparam_y} &= 
\frac{1}{\kappa \gamma} \int_{\mathcal{V}} A^i_a \mathcal{L}_{\vec{X}_{\dilparam_x, \dilparam_y}}\tilde{E}^a_i d^3x
= \frac{1}{\kappa \gamma} \left( -\dilparam_x(pc)_x 
-\dilparam_y (pc)_y - (-\dilparam_x -\dilparam_y)(pc)_z \right)\\
&= \frac{1}{\kappa \gamma}\left((pc)_z - (pc)_x\right) \dilparam_x
+ \frac{1}{\kappa \gamma}\left((pc)_z - (pc)_y\right) \dilparam_y
= \frac{-\mathbb{S}_y}{\kappa\gamma (p\mathbb{c})_y} \dilparam_x
+\frac{\mathbb{S}_x}{\kappa\gamma (p\mathbb{c})_x} \dilparam_y \\
&= \frac{-\mathbb{S}_y}{\kappa\gamma \left((pc)_y + i\alpha\vol(\vec{p})\right)} \dilparam_x
+\frac{\mathbb{S}_x}{\kappa\gamma \left((pc)_x+ i\alpha\vol(\vec{p})\right)} \dilparam_y .
\\
&= \frac{-\overline{\mathbb{S}}_y}{\kappa\gamma \left((pc)_y - i\alpha\vol(\vec{p})\right)} \dilparam_x
+\frac{\overline{\mathbb{S}}_x}{\kappa\gamma \left((pc)_x - i\alpha\vol(\vec{p})\right)} \dilparam_y .
\end{split}
\end{align}
The key point is that these generators are not constraints --- anisotropic dilatations do not approach the identity at infinity, so that they are not generated by the diffeomorphism constraint.
%which is identically zero in \mbox{Bianchi I} anyway. 
Thus, their vanishing imposes a non-trivial restriction on the physical degrees of freedom.
In fact, it is immediate from the above form that the vanishing of the above generators for all $\dilparam_x, \dilparam_y$ is equivalent to $\mathbb{S}_x \approx \mathbb{S}_y \approx 0$, 
which is equivalent to $\mathbb{S}_i \approx 0$ for all $i$ --- our classical isotropy condition.
At the same time, it is equivalent to $\overline{\mathbb{S}}_i \approx 0$. 
% Finally, that 
% $\overline{\mathbb{S}}_i$ is proportional to $\mathbb{S}_i$
% with coefficient everywhere smooth and non-vanishing --- equation (\ref{eqn:SpropSbar}) --- is consistent with the expectation that 
% $\hat{\mathbb{S}}_i \circ \iota = 0$ and 
% $\hat{\mathbb{S}}_i^\dagger \circ \iota = 0$ be equivalent
% in quantum theory.

\paragraph{Imposition of Intertwining of Signed Volume and Hubble Operator}

Imposition of canonical residual automorphism covariance and quantum isotropy has not yet uniquely determined the embedding $\iota$.  But this was expected: These conditions have only restricted the \textit{image} of $\iota$. 
As noted in \cite{behm2017}, in order to achieve uniqueness of $\iota$, one expects to impose two more conditions, such as the intertwining of two operators. The basic variables in the isotropic theory are $p$ and $c$, so it is natural to try to impose intertwining of corresponding operators with appropriate operators in the \mbox{Bianchi I} theory. One can indeed require that 
$\iota$ intertwine $\hat{p}$ with $\hat{v}^{2/3}$ in \mbox{Bianchi I}, which is equivalent to requiring that $\iota$ intertwine the signed volume $\hat{v}$ in both theories.  However, $c$ has no operator analogue in the quantum theory, but rather only exponentials of $c$ have operator analogues. Because of this, it is natural to instead require intertwining of an appropriate one-parameter family of exponentials of $c$, or operators contructed therefrom.
We choose to require intertwining of the regularized isotropic Hubble rate 
$\hat{\theta}(s)$ (\ref{Iso:dHops}) with one of the regularized directional Hubble rates
(\ref{BI:dHops}) --- specifically, we arbitrarily choose $\hat{\theta}_z(s)$ for this purpose.
With this condition imposed, we shall see that $\iota$ is uniquely determined up to an overall constant, and will then automatically intertwine $\hat{\theta}(s)$ with the other directional Hubble rates as well. Indeed, we shall see that the resulting unique $\iota$ will satisfy basically every property that could be desired from such an embedding. 

Let $\Cyl_S$ denote the space of finite linear combinations of volume eigenstates in the isotropic theory,
so that its algebraic dual, $\Cyl_S^*$, may be identified with distributional states which include possibly non-normalizable linear combinations of volume eigenstates.
% Introducing $\Cyl_S^*$ as the domain of the embedding 
% is important because (1.) physical states will be in $\Cyl_S*$
% and (2.) the adjoint of the Ashtekar-Wilson-Ewing project
% has $\Cyl_S^*$ as its natural adjoint.

\begin{theorem}
There exists an embedding $\iota$ from isotropic LQC states, $Cyl_S^*$, to \mbox{Bianchi I} quantum states, $Cyl_{BI}^*$, that (1.) is covariant under all canonical residual automorphisms, (2.) intertwines $\hat{v}$ in the two theories, and (3.) intertwines $\hat{\theta}(s)$ with $\hat{\theta}_z(s)$ for 
all $s$. This embedding is furthermore unique up to a (physically irrelevant)
overall constant, and is given explicitly by
$\iota = C \iota_0$ 
for some $C \in \mathbb{C}$.
\end{theorem}
{\startproof

%By the argument of section \ref{covsec}, 
From equation \eqref{eq:coviota}, condition (1.) imposes that $\iota$ be of the form
\begin{align*}
\iota |\Lambda \rangle = \sum_{\Lambda' \neq 0}
C(\Lambda; \Lambda') \iota_0 |\Lambda'\rangle
\end{align*}
for some $C(\Lambda; \Lambda')$. Condition (2.) then forces 
$C(\Lambda; \Lambda') = C(\Lambda) \delta_{\Lambda, \Lambda'}$ for some $C(\Lambda)$, and, finally, condition (3.) forces $C(\Lambda)$ to be a constant
$C$. 
\finishproof}

The overall constant $C$ is not a physical ambiguity, because quantum states have meaning only up to rescaling. 
Hence the embedding is physically unique, as was expected from the analysis of \cite{behm2017, behm2016}.
From now on we set $\iota$ to be equal to the embedding so selected, choosing $C=1$, so that 
\begin{align}
\label{finaliota}
\iota |\Lambda\rangle = \iota_0 |\Lambda \rangle = \sum_{\lambda_x, \lambda_y \neq 0} |\lambda_x, \lambda_y, \Lambda\rangle .
\end{align}
This is the \mbox{Bianchi I} analogue of what we have called the \textit{volume embedding} in the full theory \cite{behm2017}.

\paragraph{Remark}
In selecting the unique embedding $\iota$ above, we have required that it intertwine $\hat{\hubb}(s)$ with $\hat{\hubb}_z(s)$ for all $s$.
One can alternatively require that, for all $s$, the more basic shift operators $\deltop(s)$ intertwine with a 
slight modification of $\deltop_z(s)$, namely $\deltop'_i(s):= \widehat{\exp}\left(\frac{-ip_i c_i s}{2v}\right)$, also unitary,
for, e.g. $i=z$. 
%
% This is impossible if we instead use the operators 
% $\deltop_i(s)$ themselves: There exists no embedding which 
% intertwines $\deltop(s)$ with $\deltop_i(s)$ for all $s$ that is also 
% covariant under canonical residual diffeomorphisms, which is 
% morally equivalent to our quantum isotropy condition.
% Perhaps there is not even any embedding at all that intertwines 
% these two operators, but this is probably too strong a statement, and
% I have not checked it.
%
The resulting selected embedding is again the same. For this reason, these alternative shift operators are arguably more natural building blocks for the \mbox{Bianchi I} theory.
Indeed, one could construct a Hamiltonian constraint operator from these alternative shift operators, and the result would be equivalent to the one used here and in \cite{aw2009} when acting on states $|\vec{\lambda}\rangle$ with sufficiently large volume. We have not used this alternative simply in order to be consistent with \cite{aw2009}.

%
% Regarding the commented out "Remark 2" below:
% I haven't verified this. If its true, I think its better to state before the remark above. Correspondingly, the reminder of the condition on $\iota$ to intertwine the hubble rates for all s would be moved to this remark. In either case, we deliberately decided it is actually better to *not* mention the remark below unless a referee asks about the fact that we impose intertwining with "more than two operators."
%
%\paragraph{Remark 2:}
%If, prior to selecting $\iota$, one were to first restrict to a compatible choice from among the usual superselection sectors in both the Bianchi I and isotropic state spaces \cite{aw2009,aps2006}, then it would no longer be necessary 
%%(nor possible) 
%to impose intertwining of $\hat{\hubb}(s)$ and $\hat{\hubb}_z(s)$
%(or $\deltop(s)$ and $\deltop'_z(s)$) for \textit{all} $s$. 
%It would then suffice to impose it for only $s=s_0$, the value %appearing in the Hamiltonian constraint operator and in all of the %final quantum operators with physical significance. 
%We have opted not to do this in order to keep the derivation of $\iota$ independent of the choice of dynamics and the choice $s=s_0$ in the quantization of these operators. 

\subsection{Why the Arguments against the Volume Embedding Don't Apply in the Case of \mbox{Bianchi I}}

In the full theory paper \cite{behm2017}, we gave two arguments against the use of the volume embedding in the general case.  Here, we address each of them, and show they do not apply in the simpler case of embedding into \mbox{Bianchi I}.  First, we noted that the superposition which defines the volume embedding is in no way peaked on any geometries which are homogenious and isotropic.  Since \mbox{Bianchi I} is homogeneous, we only have to address the apparent lack of isotropy in the target of the embedding
(\ref{finaliota}).  It is clear from (\ref{finaliota}) that, for each volume eigenstate $\ket{\Lambda}$, $\embedding \ket{\Lambda}$ is a superposition of states $\ket{\lambdacoord_x, \lambdacoord_y, \Lambda}$ for which the condition $\lambdacoord_x = \lambdacoord_y = \lambdacoord_z$ is \textit{not} satisfied.  However, this condition merely describes the dimensions of the fiducial cell; it has nothing to do with the isotropy of the phase space variables $\pairparenth{ q_{ab}, K_{ab}}$.  Rather, the correct isotropy condition is the one that been the subject of this paper: That states should be annihilated by the operators (\ref{SCReQ}).  From Theorem \ref{thm:iotasymm} we know that $\embedding$ in fact does map all isotropic states into the isotropic sector of \mbox{Bianchi I}.

The second objection was that the definition of the volume embedding depends critically on the choice of basis used to define it.  However, in the present \mbox{Bianchi I} context, there is no ambiguity at all in the embedding.  As already shown above,  $\embedding$ is (up to an overall constant) the unique embedding which is covariant under canonical residual automorphisms and which intertwines the signed volume and any one of the directional Hubble rates.

\subsection{Additional Properties of the Embedding}

\begin{description}

\item[For Each $s$, $\iota$ Intertwines All of the Directional Hubble Rates $\hat{\theta}_i(s)$ with $\hat{\theta}(s)$.]
This follows from the fact that $\iota = \iota_0$, the volume embedding, and
Lemma \ref{lem:iota0int}.

\item[$\iota$ Intertwines the Hamiltonian Constraint Operators of the Isotropic and \mbox{Bianchi I} Models.]
This is immediate from the expressions (\ref{BI:Hgop}) and (\ref{SR:Hgop}) for these Hamiltonian constraint operators, 
together with the properties
$\hat{v} \circ \iota = \iota \circ \hat{v}$ 
and $\hat{\theta}_i(s) \circ \iota = \iota \circ \hat{\theta}(s)$
noted above.

\item[$\iota$ is the Adjoint of the Projector of Ashtekar and Wilson-Ewing.]
%\label{sec:adjoint}
In \cite{aw2009}, Ashtekar and Wilson-Ewing define a projector from \mbox{Bianchi I} states to isotropic LQC states given by 
\begin{align*}
\langle \Lambda | \hat{\mathbb{P}} \Psi \rangle
= (\hat{\mathbb{P}} \Psi)(\Lambda)
= \sum_{\lambda_x, \lambda_y} \Psi(\lambda_x, \lambda_y, \Lambda)
= \sum_{\lambda_x, \lambda_y} 
\langle \lambda_x, \lambda_y, \Lambda | \Psi \rangle 
\end{align*}
for all $\Psi$, so that 
\begin{align*}
\langle \Lambda | \hat{\mathbb{P}}
= \sum_{\lambda_x, \lambda_y} 
\langle \lambda_x, \lambda_y, \Lambda |
\end{align*}
and hence
\begin{align*}
\hat{\mathbb{P}}^\dagger |\Lambda \rangle
= \sum_{\lambda_x, \lambda_y} 
| \lambda_x, \lambda_y, \Lambda \rangle
= \iota |\Lambda \rangle
\end{align*}
whence $\hat{\mathbb{P}}^\dagger = \iota$.

\end{description}

\paragraph{Technical remark:}
Though $\hat{\mathbb{P}}$
maps normalizable states in the \mbox{Bianchi I} Hilbert space $\mfs{H}$ to normalizable states in the isotropic Hilbert space $\mfs{H}_S$, it is unbounded and hence only densely defined. As a consequence, its adjoint \textit{in the sense of a densely defined map $\mfs{H}_S \rightarrow \mfs{H}$} need not, and in fact does not, exist.
However, the adjoint in the algebraic dual sense always exists. The domain of 
$\hat{\mathbb{P}}$ can be taken to be, for example, $\Cyl_{BI}$;
with this choice, its range is $\Cyl_S$. 
The adjoint in the algebraic dual sense, $\hat{\mathbb{P}}^\dagger: \Cyl_S^* \rightarrow \Cyl_{BI}^*$ can then be restricted to a map $\hat{\mathbb{P}}^\dagger: \mfs{H}_S \rightarrow \Cyl_{BI}^*$. This is the map which equals our selected embedding $\iota$ up to constant rescaling, mapping all non-zero states in $\mfs{H}_S$ into non-normalizable states in $\Cyl_{BI}^*$.

\section{Origins of the Embedding Properties}
\label{sec:origins}

In contrast to what is expected in the full theory \cite{behm2017}, we have seen above that, for the embedding into \mbox{Bianchi I}, the following holds:
\begin{enumerate}
\item 
Not only is it possible to impose the quantization of the symmetry conditions (\ref{SR:scred}) 
consistently in the quantum theory, 
but also possible to simultaneously impose their adjoint.  
Furthermore, a natural embedding of the quantum isotropic model into the common kernel of the quantum conditions and their adjoint exists. 
\item Every operator of interest preserves the image of this natural embedding, and so is intertwined with some operator on $\Hil_S$ which turns out to be exactly the corresponding operator in the isotropic model.
\end{enumerate}
These are surprisingly strong results. 
The first result implies that the quantization 
of the real and imaginary parts
of $\mathbb{S}_i$ --- $\widehat{{\rm Re} \mathbb{S}}_i := \frac{1}{2}(\mathbb{S}_i + \mathbb{S}_i^\dagger)$ and 
$\widehat{{\rm Im}\mathbb{S}}_i := \frac{1}{2i}(\mathbb{S}_i - \mathbb{S}_i^\dagger)$ --- each annihilate the image of
$\iota$.
As first argued by Dirac \cite{dirac1964}, 
it is physically correct to impose a given system of real
constraints strongly in quantum theory only if it 
forms a \textit{first class system}.  Do ${\rm Re} \hat{\mathbb{S}}_i$ and ${\rm Im} \hat{\mathbb{S}}_i$ form such a system? 
\textit{Indeed they do.}
This is equivalent to none other than the Poisson bracket (\ref{eqn:SSbarfirst}).
%which holds in \mbox{Bianchi I}, but not the full theory.

In the following, we will see that the second result is likewise foreshadowed by classical Poisson brackets that indicate that, in fact, one expects the second result to be true for \textit{every operator invariant under proper anisotropic dilatations}, 
and thus in particular for every operator invariant under residual automorphisms.
Finally, we note that the Poisson brackets foreshadowing both of the above results hold thanks to the fact that $\overline{\mathbb{S}}$ is proportional to $\mathbb{S}$ with coefficient smooth and non-vanishing everywhere, and trace the source of this to an observation about the physics of the \mbox{Bianchi I} phase space.

\subsection{Poisson brackets indicating that $\iota$ should intertwine all proper-dilatation invariant operators}
\label{ss:genpbresult}

 We here prove that any function $F$ on the \mbox{Bianchi I} phase space invariant under proper anisotropic dilatations --- and hence in particular any $F$ invariant under residual automorphisms --- satisfies 
\begin{align}
\label{eq:PBgen}
\{ F, \mathbb{S}_i \} = \sum_j \lambda_i{}^j \mathbb{S}_j
\end{align}
for some matrix of phase space functions $\lambda_i{}^j$.
This leads to the expectation that an appropriate quantization of each such quantity will preserve the quantum isotropic sector, hence preserve the image of the embedding, and therefore be intertwined with corresponding operators in the isotropic theory, 
a fact which we have already seen is true for $F$
equal to the volume of the fiducial cell, 
the directional Hubble rates, and the Hamiltonian constraint operators.\footnote{The 
full theory paper \cite{behm2017} also includes a notion of average spatial curvature which is invariant under gauge and diffeomorphisms.  
This is identically zero in the present \mbox{Bianchi I} framework and so is also (trivially) intertwined here.}
%Note, For the specific cases of 
%the Hamiltonian constraint, volume, and directional 
%Hubble rates, the equality also hold also for $\vol(\vec{p}) = 0$.

Let us begin with a general argument that the analogue of (\ref{eq:PBgen}) in the full theory \textit{almost} holds.
This will allow us to see precisely the special property of the 
\mbox{Bianchi I} phase space that enables the argument to be completed.
Suppose we are given a function $F$ on the full theory phase space $\Gamma_\Full$ which is invariant under all spatial diffeomorphisms and 
local gauge rotations -- that is, invariant under all automorphisms of the $SU(2)$ principal fiber bundle.
Let $\overline{\Gamma_\Full}$ be the bundle-automorphism-covariant homogeneous isotropic sector, defined as the set of points $\eta \in \Gamma_\Full$
such that $\mathbb{S}[f,g](\eta) = 0$ for all $f,g$.
From \cite{behm2017},$\eta \in \overline{\Gamma_\Full}$ if and only if, for one of the three homogeneous-isotropic symmetry groups $\mathcal{G}$ (Euclidean group, $SO(4)$, or $SO(3,1)$), there exists
\textit{some} action $\rho$ of $\mathcal{G}$, via bundle automorphisms, such that 
$\rho(\alpha)\eta = \eta$ for all $\alpha \in \mathcal{G}$.
Because $F$ is automorphism invariant, its Hamiltonian flow cannot map one out of the symmetric sector $\overline{\Gamma_\Full}$. Heuristically,
one can see this because, in order for the flow of $F$ to map a point in $\overline{\Gamma_\Full}$ out of itself, $F$ would need to determine `where' the inhomogeneity
or `in which direction' the anisotropy arises.  But because $F$ is invariant under diffeomorphisms and gauge rotations, this is not possible.
More explicitly, let $\eta \in \overline{\Gamma_\Full}$ be given, let $\mathcal{G}$ be the corresponding homogeneous-isotropic symmetry group, and let $\rho$ be the corresponding action of $\mathcal{G}$.
Let $\Phi_F^t: \Gamma_\Full \rightarrow \Gamma_\Full$ denote the Hamiltonian flow generated by $F$ on $\Gamma_\Full$.
Because both $F$ and the Poisson brackets on $\Gamma_\Full$ are automorphism covariant, so is $\Phi_F^t$ for each $t$, so 
that $\varphi \circ \Phi_F^t = \Phi_F^t \circ \varphi$ for all automorphisms $\varphi$ and all $t \in \mathbb{R}$. Thus, in particular, 
for all $\alpha \in \mathcal{G}$ and all $t \in \mathbb{R}$, we have
\begin{align*}
\rho(\alpha) \Phi_F^t (\eta) = \Phi_F^t ( \rho(\alpha) \eta) = \Phi_F^t(\eta)
\end{align*}
so that $\Phi_F^t(\eta) \in \overline{\Gamma_\Full}$ as well. 
 Thus $\mathbb{S}[f,g](\Phi_F^t(\eta)) = 0$ for all $t$.  Taking the derivative with respect to $t$
and setting $t$ to zero yields
\begin{align*}
\{ F, \mathbb{S}[f,g] \}(\eta) = 0
\end{align*}
for all $f,g$, and all $\eta \in \overline{\Gamma_\Full}$. 
As $\overline{\Gamma_\Full}$ is the zero set of $\mathbb{S}[f,g]$ for all $f,g$, and since the topology of 
$\Gamma_\Full$ is trivial, it follows that 
\begin{align}
\label{eq:fullgen}
\{F, \mathbb{S}[f,g] \} = \mathbb{S}[h,k] + \overline{\mathbb{S}[\tilde{h}, \tilde{k}]}
\end{align}
for some $h,k,\tilde{h}, \tilde{k}$ depending on $f$ and $g$ and possibly the phase space point.

The above argument goes through also for the \mbox{Bianchi I} case, with minimal modification. Let $\Gamma$ denote the \mbox{Bianchi I} phase space
as in section \ref{s:BIrev}.
The only modifications required to adapt the above argument to this case are the following:
\begin{enumerate}
\item The full group of bundle automorphisms is replaced by the canonical residual automorphisms.
${\rm Aut}^C_R$.
\item 
Instead of three possibilities for the symmetry group $\mathcal{G}$, there is only one, namely the residual rotation group, ${\rm Rot}_R$ --- that part of the Euclidean group with well-defined and non-trivial action in the Bianchi I context.
%
%the reduction of the Euclidean group in the Bianchi I context 
%
% Since space is flat in Bianchi I, the only possibility for the maximal symmetry group $\mathcal{G}$ is the Euclidean group, which, in the Bianchi I context, reduces to the residual rotation group, ${\rm Rot}_R$.
%
%  Instead of three possibilities for the maximal symmetry group $\mathcal{G}$, there is only one, namely the residual rotation group, ${\rm Rot}_R$.  
\end{enumerate}
That is, in the \mbox{Bianchi I} case
one need only require that $F$, now a function on $\Gamma$, be invariant under ${\rm Aut}^C_R$. 
Additionally, for all $\eta \in \Gamma$, by Theorem \ref{thm:sym},
 $\mathbb{S}_i(\eta) = 0$ (i.e., $\eta \in \overline{\Gamma}$) 
 if and only if there exists some proper, and hence canonical, residual automorphism
$\varphi$ such that $\eta$ is invariant under the action 
$\rho(\pi):= \varphi \circ \pi \circ \varphi^{-1}$ of all $\pi \in {\rm Rot}_R$.
This, combined with the invariance of $F$ and the Poisson brackets
under ${\rm Aut}^C_R$, allows the above argument in the full theory
to be repeated unchanged in the \mbox{Bianchi I} case. \textit{Thus 
(\ref{eq:fullgen}) holds also in the \mbox{Bianchi I} case,} where it is more 
conveniently written as
\begin{align}
\label{eq:BIgen}
\{F, \mathbb{S}_i \} = \sum_j \left(h_i{}^j\mathbb{S}_j 
+ k_i{}^j\overline{\mathbb{S}}_j \right)
\end{align}
for some possibly phase space dependent $h_i{}^j$, $k_i{}^j$.
This is so far exactly analogous to the full theory.
\textit{What is special} in the \mbox{Bianchi I} case is equation 
(\ref{eqn:SpropSbar}), which allows
(\ref{eq:BIgen}) to be rewritten precisely in the form (\ref{eq:PBgen}) claimed.
Furthermore, equation (\ref{eq:BIgen}) at any given phase space point $\eta$ depends on $F$ only in a neighborhood of $\eta$.
As a consequence, the invariance of $F$ under the full group
of residual canonical automorphisms is not relevant for the validity of (\ref{eq:BIgen}), 
but only invariance under the identity component of this group, 
namely the canonical anisotropic dilatations.
That is, it is actually sufficient for $F$ to be invariant under the smaller group of canonical anisotropic dilatations
for (\ref{eq:BIgen}) to hold. 
The volume of the fiducial cell, the directional Hubble rates, 
and the Hamiltonian constraint are all examples of such $F$'s.

\paragraph{Explicit Calculation in Cases of Interest}

We here explicitly calculate the matrix of phase space functions 
$\lambda^i{}_j$ in \eqref{eq:PBgen} for the cases of $F$
corresponding to the operators already shown to be intertwined by the embedding $\iota$.
We do this both for concreteness,
as well as to perform a check on the general arguments above.
% We do this both because these cases are the most directly relevant for explaining these intertwining results,
% as well as to perform a check on the general arguments above.

\paragraph{\bf The volume of the fiducial cell.}
From the expressions (\ref{BI:aofp}), (\ref{BI:Bdef}), (\ref{SR:scred}), and $\{c^i, p_j\} = \kappa \gamma \delta^i_j$, one calculates
\begin{align}
\label{eq:SCVPB2}
\{ \mathbb{S}^i, \vol(\vec{p}) \} = \frac{\kappa\gamma}{2}
\frac{1}{\gamma \theta_i + i\alpha} \mathbb{S}^i .
\end{align}
% so that, for $F = \vol(\vec{p})$,
% $\lambda_i{}^j = \frac{\kappa\gamma}{2}
% \frac{1}{\gamma \theta_i + i\alpha} \delta_i^j$.

\paragraph{\bf The directional Hubble rates.}
Similarly, from the definition (\ref{BI:dHub}), 
\begin{align}
\{\mathbb{S}^i, \hubb_k \}
= -\frac{\kappa}{2 \vol(\vec{p})} \frac{\gamma \theta_k + i \alpha}{\gamma \theta_i + i\alpha}
\mathbb{S}^i .
\end{align}
% so that for $F = \hubb_k$, $\lambda_i{}^j = -\frac{\kappa}{2 \vol(\vec{p})} \frac{\gamma \theta_k + i \alpha}{\gamma \theta_i + i\alpha}\delta_i^j$ .

\paragraph{\bf The Hamiltonian constraint.}
From equation (\ref{BI:Hred}), using $N = \vol(\vec{p})^n$ 
from equation (\ref{BI:Hgop}), and using the above two Poisson brackets, we have
\begin{align}
\label{eq:SCBPB}
\begin{split}
\{ \mathbb{S}^i, H_g \}
&= \frac{\vol(\vec{p})^n}{2(\gamma \theta_i + i\alpha)}
\left(\frac{1-n}{\gamma} \sum_{j<k} \theta_j \theta_k
+ 2i\alpha \sum_{j=1}^3 \theta_j \right)\mathbb{S}^i\\
&= \frac{1}{\gamma \theta_i + i\alpha}
\left( \frac{\gamma (n-1)}{2\vol(\vec{p})} H_g
+ i\alpha \vol(\vec{p})^n \sum_{j=1}^3 \theta_j \right)\mathbb{S}^i .
\end{split}
\end{align}
% whence, for $F=H_g$,
% $\lambda_i{}^j$ is given by the above coefficient of $\mathbb{S}^i$ times $\delta^i_j$.

\subsection{Deeper Source of Surprising Simplifications in the \mbox{Bianchi I} Case}

At the start of this section, we have summarized a number of surprisingly congruous features of the quantum isotropic symmetric sector of \mbox{Bianchi I} and a natural embedding of the quantum isotropic theory into it. In section \ref{ss:genpbresult} we have exhibited reason to expect that these features extend even further.
Furthermore, the argument above in section \ref{ss:genpbresult}, 
as well as the arguments in section
\ref{SR:Constraints}, show that all of these unexpected results, in the end, can be traced to  
the fact (\ref{eqn:SpropSbar}), that
\textit{$\overline{\mathbb{S}_i}$ and $\mathbb{S}_i$ are proportional to each other with coefficient everywhere smooth and non-vanishing}. 
Why does this property hold specifically in \mbox{Bianchi I}? This property is directly implied by the fact that the real part of the symmetry condition is proportional to the imaginary part by an everywhere smooth and real coefficient:
\begin{align}
    {\rm Re \mathbb{S}_i} = \left(\frac{\gamma \theta_i}{\alpha}\right) {\rm Im \mathbb{S}_i} .
\end{align}
This coefficient is non-vanishing throughout $\Gamma$ except where $c_i = 0$. This proportionality is a reflection of the fact that
the real and imaginary parts of the symmetry conditions $\mathbb{S}_i \approx 0$ are not independent, but rather \textit{the imaginary part implies the real part, and almost vice versa}. 

Why are only half of the symmetry conditions independent? To see the answer to this question, we note that 
the fact that the spin-connection is flat means that the spatial geometry is \textit{unique up to diffeomorphism} in \mbox{Bianchi I} 
--- i.e., the triad $\tilde{E}^a_i$ by itself has \textit{no diffeomorphism and gauge invariant information}. 
This can also be seen more directly. Consider the action \eqref{BI:anis},  
\eqref{BI:pref} of the residual diffeomorphisms in the \mbox{Bianchi I} case.  It is easy to see that this action acts transitively on the space of all 
non-degenerate densitized triads $\tilde{E}^\mathbf{a}_i = p_i \delta^\mathbf{a}_i$ in \mbox{Bianchi I}.  The same is also true for the space of all 
connections $A^i_\mathbf{a} = c^i \delta^i_\mathbf{a}$ if one restricts to connections with no vanishing components.  Thus, $\tilde{E}^a_i$ by itself and 
$A^i_a$ by itself (basically) each contain no diffeomorphism invariant information. 
\textit{Only the relation between them contains diffeomorphism invariant
information}. Because the symmetry condition $\mathbb{S}_i \approx 0$ is diffeomorphism invariant, 
this means that it implies no condition on $\tilde{E}^a_i$ or $A^i_a$ separately, but only a condition on their relation to each other.
Thus, if the residual diffeomorphism freedom is used to completely fix $\tilde{E}^a_i$ arbitrarily, the symmetry condition yields a condition on $A^i_a$ only, or vice-versa, so that effectively the symmetry 
constraint is a constraint on only ``half'' of the variables. 

This last observation also resolves a tension in the fact that, as mentioned above, 
the set of constraint functions $\{{\rm Re}\mathbb{S}_i,
{\rm Im}\mathbb{S}_i\}$ are first class. 
This set imposes the diffeomorphism invariant part of 
the symmetry condition on both $\tilde{E}^a_i$ and $A^i_a$, conjugate variables. Real-valued constraint functions imposing symmetry on conjugate variables 
normally would form a \textit{second} class set, not a first class set \cite{engle2006, behm2017, behm2016}.
However, as noted in the last paragraph, because our symmetry conditions impose only the \textit{diffeomorphism invariant part} of 
homogeneity and isotropy, in the present \mbox{Bianchi I} case, the conditions impose no conditions on either $\tilde{E}^a_i$ or $A^i_a$
separately, but only on the relation between the two.  Thus, specifically in this \mbox{Bianchi I} case, \textit{no symmetry condition is imposed 
separately and simultaneously on any conjugate components of variables}, so that the usual argument leading to the conclusion that the real and imaginary parts of the constraint functions should be second class does not apply.

\section{Discussion}

In the work \cite{behm2016, behm2017}, we introduced 
a gauge- and diffeomorphism-invariant --- that is, principal-bundle-automorphism invariant --- notion of homogeneous and isotropic states in full loop quantum gravity, together with a strategy for constructing an embedding of loop quantum cosmology states into the space of such full theory states. We proposed that the resulting embedding be used to relate proposals for dynamics in full LQG with choices of dynamics in LQC, where observational consequences can be more easily calculated.

In the present paper, as a test, we have applied these ideas to the simpler case of embedding into \mbox{Bianchi I}, with surprising success. In this simpler context, 
the automorphism-invariant conditions for homogeneity and isotropy reduce to residual-automorphism-invariant conditions $\mathbb{S}^i \approx 0$ for isotropy. They can be easily quantized in the manner analogous to that suggested for the full theory in \cite{behm2016, behm2017} and using the methods of \cite{aw2009}, yielding 
operators $\hat{\mathbb{S}}^i$. 
These operators are non-hermitian, and may be thought of as the ``holomorphic part'' of the symmetry conditions in the Gupta-Bleuler sense.

Furthermore, we have shown that there exists a \textit{unique} embedding, of isotropic LQC into \mbox{Bianchi I} states, satisfying the following three conditions:
\begin{enumerate}
\item It be covariant under all residual automorphisms with well-defined actions on quantum states --- the canonical residual automorphisms.
\item It intertwine the signed volume operator in the two models.
\item It intertwine the regularized directional Hubble rate $\hat{\theta}_z(s)$ in the \mbox{Bianchi I} model with the Hubble rate $\hat{\theta}(s)$ in the isotropic model for all $s$.
\end{enumerate}
The embedding $\iota$ so selected then automatically satisfies the following further properties:
%
% I thought it best to make the conclusions below bulleted, and the assumptions above enumerated, or vice versa, to distinguish them better. 
%
\begin{itemize}
\item It is annihilated by the quantum isotropy conditions 
$\hat{\mathbb{S}}_i$ --- that is, it is an embedding into the sector of quantum isotropy.
\item It intertwines \textit{all} of the directional Hubble rates $\hat{\theta}_i$ with $\hat{\theta}$.
\item It intertwines the Hamiltonian constraint operators in the isotropic and \mbox{Bianchi I} models. 
\item It is the adjoint of the projector from \mbox{Bianchi I} states to isotropic states introduced by Ashtekar and Wilson-Ewing in \cite{aw2009}.
\end{itemize}
In particular, $\iota$ intertwines every operator of interest in the isotropic and \mbox{Bianchi I} models. From classical analysis, we in fact have seen that we expect \textit{all} canonical residual automorphism invariant operators in the \mbox{Bianchi I} and isotropic models, if appropriately quantized, to be intertwined by $\iota$.
Equally surprisingly, and perhaps at the root of this, we have seen that $\iota$ is not only annihilated by $\hat{\mathbb{S}}_i$, \textit{but also by the adjoints} $\hat{\mathbb{S}}_i^\dagger$ --- by \textit{both} the ``holomorphic'' and ``anti-holomorphic'' parts of the symmetry conditions. 
In section 5, we traced these last two surprising results to the fact that, in Bianchi I, 
${\rm Re}\mathbb{S}_i$ is proportional to ${\rm Im}\mathbb{S}_i$
with coefficient everywhere finite and smooth, a fact which does not hold in the full theory \cite{behm2017}.
Though, in the full theory, we thus expect the obvious interesting operators to not preserve the quantum homogeneous isotropic sector, nevertheless, in this same work \cite{behm2017} we have laid out a strategy to handle the expected resulting added complication in this case.
% Though, as noted in \cite{behm2017}, we expect the obvious interesting operators in the full theory to not preserve the quantum homogeneous isotropic sector, nevertheless, in this same work we have laid out a strategy to handle the expected resulting added complication in the full theory case.

\section*{Acknowledgements}

The authors are grateful to Ted Jacobson, Atousa Chaharsough Shirazi, Brajesh Gupt, Jorge Pullin, Parampreet Singh, Xuping Wang, and Shawn Wilder for discussions, and to Edward Wilson-Ewing for pointing out a sign error in \cite{aw2009} that was corrected in \cite{wilsonewingthesis}.
This work was supported in part by NSF Grants PHY-1205968, PHY-1505490, and PHY-1806290, and by NASA through the University
of Central Florida's NASA-Florida Space Grant Consortium. 

\vspace{1.5em}

%%%\bibliography{..\bibtex_files\jonsbib}
%\bibliography{jonsbib}{}
%\bibliographystyle{ieeetr}
%%% Reason for ieeetr: (1.) bibliography is in order of citation,
%%% (2.) titles are given for articles, but (3.) no full author first names
%%% are given, only initials, and (4.) format for authors is same for
%%% both articles and books
%%%
%
%\end{document}

\end{document}